\documentclass[sigconf]{acmart}

\usepackage{amsmath}
\usepackage{amsthm}
\usepackage{bm}
\usepackage{array}
\usepackage{makecell}
\usepackage{multirow}
\usepackage{tabularx}
\usepackage{paralist}
\usepackage{graphicx}
\usepackage{caption}
\usepackage{subcaption}
\usepackage{url}
\usepackage{xcolor}
\usepackage{enumitem}
\usepackage{float}
\usepackage[flushleft]{threeparttable}
\usepackage[ruled,vlined,linesnumbered]{algorithm2e}
 
\usepackage{booktabs}
\usepackage{amssymb}
\usepackage{amsmath} 
\usepackage{courier}
\usepackage{enumitem}
\usepackage{bbding}
\usepackage{pifont}
\usepackage{setspace}
\usepackage{array,multirow,multicol}
\usepackage{caption, subcaption}
\usepackage{algpseudocode} 
\usepackage{lipsum}

\newcommand{\PreserveBackslash}[1]{\let\temp=\\#1\let\\=\temp}
\newcolumntype{C}[1]{>{\PreserveBackslash\centering}p{#1}}
\newcolumntype{R}[1]{>{\PreserveBackslash\raggedleft}p{#1}}
\newcolumntype{L}[1]{>{\PreserveBackslash\raggedright}p{#1}}
\newcommand{\tabincell}[2]{\begin{tabular}{@{}#1@{}}#2\end{tabular}}  

\newif{\ifhidecomments}
\ifhidecomments
    \newcommand{\viv}[1]{}
\else
    \newcommand{\viv}[1]{\textcolor{red}{[#1 ---\textsc{viv}]}}
\fi

\DeclareMathOperator{\Tr}{Tr}
\AtBeginDocument{%
  \providecommand\BibTeX{{%
    \normalfont B\kern-0.5em{\scshape i\kern-0.25em b}\kern-0.8em\TeX}}}

\copyrightyear{2024}
\acmYear{2024}
\setcopyright{acmlicensed}\acmConference[WSDM '24]{Proceedings of the 17th
ACM International Conference on Web Search and Data Mining}{March 4--8,
2024}{Merida, Mexico}
\acmBooktitle{Proceedings of the 17th ACM International Conference on Web
Search and Data Mining (WSDM '24), March 4--8, 2024, Merida, Mexico}
\acmPrice{15.00}
\acmDOI{10.1145/3616855.3635832}
\acmISBN{979-8-4007-0371-3/24/03}

\begin{document}

\title{Towards Mitigating Dimensional
Collapse of Representations in Collaborative Filtering}

\author{Huiyuan Chen}
\email{hchen@visa.com}
\affiliation{%
  \institution{Visa Research}
  \city{Palo Alto}
  \country{USA}}

\author{Vivian Lai}
\email{viv.lai@visa.com}
\affiliation{%
  \institution{Visa Research}
  \city{Palo Alto}
  \country{USA}}

\author{Hongye Jin}
\email{jhy0410@tamu.edu}
\affiliation{%
  \institution{Texas A\&M Unversity}
  \city{College Station}
  \country{USA}}

\author{Zhimeng Jiang}
\email{zhimengj@tamu.edu}
\affiliation{%
  \institution{Texas A\&M Unversity}
  \city{College Station}
  \country{USA}}

\author{Mahashweta Das}
\email{mahdas@visa.com}
\affiliation{%
  \institution{Visa Research}
  \city{Palo Alto}
  \country{USA}}

\author{Xia Hu}
\email{xia.hu@rice.edu}
\affiliation{%
  \institution{Rice University}
  \city{Houston}
  \country{USA}}

\renewcommand{\shortauthors}{Huiyuan Chen et al.}

\begin{abstract}

Contrastive Learning (CL) has shown promising performance in collaborative filtering. The key idea is to generate augmentation-invariant embeddings by maximizing the Mutual Information between different augmented views of the same instance. 
However, we empirically observe that existing CL models suffer from the \textsl{dimensional collapse} issue,  where user/item embeddings only span a low-dimension subspace of the entire feature space. This suppresses other dimensional information and    weakens the distinguishability of embeddings. 
Here we propose a non-contrastive learning objective, named nCL, which explicitly mitigates dimensional collapse of representations in  
collaborative filtering. Our nCL aims to achieve geometric properties of \textsl{Alignment} and \textsl{Compactness} on the embedding space. In particular, the alignment tries to push together representations
of positive-related user-item pairs, while compactness tends to find the optimal coding length of user/item embeddings, subject to a given distortion. More importantly, our nCL does not require data augmentation nor negative sampling during training, making it scalable to large datasets. Experimental results demonstrate the superiority of our nCL.


\end{abstract}


\maketitle

\section{Introduction}


Collaborative Filtering  is a prominent technique~\cite{mao2021simplex,he2017neural},  which considers the users' historical contexts (\textsl{e.g.}, purchases and clicks) and assumes that users who share similar preferences  tend to make similar decisions. Collaborative filtering algorithms have evolved over the years, from simple Factorization Machines~\cite{koren2009matrix} to deep neural networks, such as Neural Matrix Factorization~\cite{he2017neural}  and Graph Neural Networks~\cite{he2020lightgcn,mao2021ultragcn,chen2021structured,yan2023from}. While collaborative models have achieved enormous success in industries, they are often  bottlenecked by  limited annotated labels, which highly sacrifices the performance for long-tail or cold-start items~\cite{wu2021self,zhou2020s3,chen2022graph,liang2023learn,lai2023enhancing}. 

Recently, Contrastive Learning (CL) has emerged as a promising paradigm for learning  representations without costly labels \cite{wang2020understanding, chen2020simple, gao2021simcse}. Contrastive methods operate on the assumption that the learned representations should be invariant to certain input transformations, rather than relying on task-specific supervision.  Several studies have sought to harness the idea of CL in recommendation ~\cite{wu2021self,yu2022graph,lin2022improving,zhou2021contrastive}. For example, SGL~\cite{wu2021self} generates node embeddings by maximizing
the Mutual Information  among different augmented views of the same node. On the other hand, SimGCL \cite{yu2022graph} enhances recommendation accuracy and training efficiency by introducing uniform noises to the latent embedding space to create contrastive views. 
The simplicity and effectiveness of CL offer a promising paradigm for addressing the cold-start issue \cite{wu2021self} and reducing exposure bias in recommender systems \cite{zhou2021contrastive}.

\begin{figure}[t!]
\centering
\includegraphics[width=7.4cm]{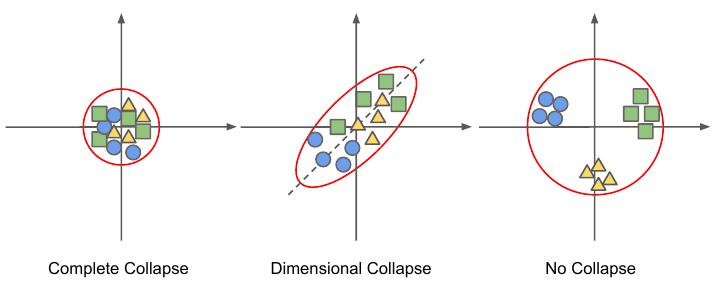}
\caption{Different collapsing patterns (complete collapse, dimensional collapse, and no collapse) for three item groups. }
\label{fig:graph}
 \vspace{-3mm}
\end{figure}

However, one of the major challenges in contrastive  learning is the neural collapse problem~\cite{wang2020understanding, hua2021feature, he2022exploring}.  Figure \ref{fig:graph} illustrates different collapse phenomena in the low-dimensional space when given three groups of items: 1) \textsl{Complete collapse}, where the feature embeddings of all
items tend to collapse to a trivial representation; 2) \textsl{Dimensional Collapse}, where the item embeddings only span a low-dimension subspace of the entire feature space; 3) \textsl{No Collapse}, where representations of items in the same group are close to each other, while different groups of items are as far from each other as possible. This is the desired scenario for representation learning~\cite{wang2020understanding,hua2021feature,he2022exploring}. In practice, the issue of complete collapse  can be carefully addressed by different contrastive mechanisms, such as using InfoNCE loss~\cite{wu2021self}, cosine contrastive loss~\cite{mao2021simplex}, or adding uniform perturbations~\cite{yu2022graph}.  However, very few efforts have been devoted to investigating the dimensional collapse of user and item representations in collaborative filtering. 


In this work, we have empirically observed that many existing contrastive models, such as SGL~\cite{wu2021self}, suffer from the issue of dimensional collapse, which results in a rank deficient embedding space. To address this issue,  we propose a non-contrastive  learning objective, named nCL, which explicitly mitigates dimensional collapse of  representations in  
collaborative filtering. More specifically, our nCL aims to achieve two geometric properties on the embedding hypersphere: 1) \textsl{Alignment}, which pushes together representations of positive-related user-item pairs; 2) \textsl{Compactness}, which  determines the optimal coding length of user/item embeddings, inspired by recent progress in coding rate reduction~\cite{yu2020learning,chan2022redunet, baek2022efficient}.  Additionally, we seamlessly combine clustering and representation learning to learn  diverse and discriminative  user/item embeddings. Experimental results on four public datasets demonstrate that nCL outperforms state-of-the-art contrastive learning methods. We summarize our main contributions as follows:
\begin{itemize}[leftmargin=*]
    \item We empirically show that existing contrastive GNN-based models suffer from dimensional collapse whereby all the user/item embedding vectors fall into a lower-dimensional subspace, rather than the entire available space. This weakens the distinguishability of  representations in    recommendation. 
    \item We propose a simple but effective learning
objective, named nCL, that directly  mitigates dimensional collapse of  representations by achieving two geometric properties on the hypersphere: alignment and compactness.
\item We show that our  nCL does not require complex  data augmentation nor negative sampling as in  many  baselines. This allows for small batch size training and makes nCL scalable to large datasets. Extensive experiments on   public datasets show that  our nCL outperforms state-of-the-art contrastive methods.
\end{itemize}

\section{Related Work}

 \subsection{\textbf{Collaborative Filtering}}
Collaborative filtering  aims to learn sophisticated feature interactions between users and items ~\cite{rendle2009bpr}. Matrix factorization is an early approach to learn the latent embeddings of users and items   and use the inner product to predict the users' preference~\cite{koren2009matrix}. Motivated by the success of deep neural networks~\cite{he2017neural,guo2017deepfm,chen2022denoising,song2023,wang2022improving,xu2023kernel,yeh2022embedding}, deep recommender systems can be further used  to exploit more complex and nonlinear feature interactions between users and items. Some popular models include NCF~\cite{he2017neural}, DeepFM~\cite{guo2017deepfm}, DIN~\cite{zhou2018deep}, etc. 

  Recently, Graph Neural Networks  explore the implicit high-order proximity between nodes in the bipartite graphs, which is helpful for discovering deeper connections between users and items~\cite{wang2019neural,monti2017geometric,liang2022reasoning,chen2023sharpness}. For example,  
    NGCF~\cite{wang2019neural} proposes an embedding propagation layer to harvest the high-order collaborative signals. LightGCN~\cite{he2020lightgcn}  simplifies the design of NGCF by discarding the nonlinear transformation.  UltraGCN~\cite{mao2021ultragcn} further  skips infinite layers of message passing for efficient recommendation. Although encouraging performance has been achieved,  GNN-based methods are often limited by  limited annotated labels~\cite{wu2021self,zhang2023mixupexplainer,luo2021learning,luo2020parameterized}.  For example, the low-degree users/items only have very few neighbors, making it insufficient to learn good representations of users/items in the message-passing schema under GNNs. 


\subsection{\textbf{Contrastive Learning}}
Contrastive Learning (CL) has been  widely adopted in various applications such as computer vision~\cite{chen2020simple}, language modeling~\cite{gao2021simcse}, graph data mining~\cite{you2020graph}. Recently, many researchers have attempted to leverage the idea of CL in recommendation,  especially Graph Contrastive Learning  ~\cite{wu2021self,yu2022graph,lin2022improving,liang2023knowledge,zhou2021contrastive}. The key idea is to let the neural networks learn from the intrinsic structure of the raw data. For example, SGL~\cite{wu2021self} is an early work to generate node embeddings by maximizing
the Mutual Information between different augmented views of the same node. SimGCL~\cite{yu2022graph} introduces uniform noises to the latent embedding space to create contrastive views. NCL~\cite{lin2022improving}   explicitly incorporates the potential neighbors into the prototype-contrastive objective. 
Meanwhile, S$^3$-Rec~\cite{zhou2020s3} utilizes the intrinsic data correlation to derive self-supervision signals and enhance the data representations via pre-training methods for improving sequential recommendation with a large margin.

However, contrastive methods usually suffer from the dimensional collapse if the models are not well regularized~\cite{jing2021understanding}. The concentration of information on partial dimensions weakens the distinguishability of representations for downstream tasks. One prominent direction is to send a subset of representations to the loss function, \textsl{i.e.,} DirectCLR~\cite{jing2021understanding} directly optimizes the representation space of images without relying on an explicit trainable projector. Nevertheless, this method does not hold for \textsl{non-i.i.d} data (\textsl{e.g.}, graph) as the subset of node representations may be well-correlated due to the graph convolutions. Recently, DirectAU~\cite{wang2022towards} proposes a new learning objective that directly optimizes the properties of alignment and uniformity of user/item representations, which has achieved good results for recommendation. In contrast, we aim to propose a promising  paradigm for achieving a good balance among alignment, uniformity, and making full use of dimensional space, which leads to  high-quality embeddings for recommendation.

\section{Problem and Background}

\subsection{Problem Setup}

In this work,  we mainly focus on the task of the implicit feedback recommendation (\textsl{e.g.}, click, view, etc.), where the behavior data contains a set of users $\mathcal{U}=\{u\}$, a set of items $\mathcal{I}=\{i\}$, and a set of observed user-item interactions $\mathbf{R} \in \mathbb{R}^{|\mathcal{U}| \times  |\mathcal{I}|}$, where $|\mathcal{U}|$ and $|\mathcal{I}|$ denote the number of users and items, respectively, and $\mathbf{R}_{ui} = 1$ if the user $u$ has interacted with the item $i$ before, $0$ otherwise. 

The goal of collaborative filtering is to estimate the preference score $\hat{y}_{ui}  \in \mathbb{R}$ for each unobserved user-item pair $\{(u,i)|\mathbf{R}_{ui} = 0\}$, indicating how likely the user $u$ prefers the item $i$. We next revisit several popular graph-based recommenders, including LightGCN~\cite{he2020lightgcn}, SGL~\cite{wu2021self}, and DirectAU~\cite{wang2022towards}.

\subsection{LightGCN}

LightGCN~\cite{he2020lightgcn} aims to update the representation of each user/item by aggregating messages from its neighbors:
\begin{equation*}
	\begin{aligned}
	\label{eq1}
	\mathbf{e}_{u}^{(l+1)} &=\sum_{i \in N_{u}} \frac{1}{\sqrt{\left|\mathcal{N}_{u}\right| \left|\mathcal{N}_{i}\right|}} \mathbf{e}_{i}^{(l)}, \quad \mathbf{e}_{i}^{(l+1)} =\sum_{u \in \mathcal{N}_{i}} \frac{1}{\sqrt{\left|\mathcal{N}_{i}\right| \left|\mathcal{N}_{u}\right|}} \mathbf{e}_u^{(l)},
	\end{aligned}
\end{equation*}
where   $\mathbf{e}_{u}^{(l)}, \mathbf{e}_{i}^{(l)} \in \mathbb{R}^d$ denote the user and item embeddings in the $l$-layer, respectively; $d$ is the embedding size;    $\mathcal{N}_{u}$ ($\mathcal{N}_{i}$) denotes the set of neighbors connected to user $u$ (item $i$). After $L$ layers, LightGCN adopts a weighted sum pooling function to obtain the final embedding for each user $u$ or item $i$, \textsl{i.e.}, $\mathbf{e}_u = f_\text{pool}(\{\mathbf{e}^{(l)}_u, 0 \le l \le L\})$.   Finally, one can  predict the preference score for $(u,i)$ by using inner product: $\hat{y}_{ui} = \mathbf{e}_u^\top\mathbf{e}_i$.

  LightGCN employs the pairwise Bayesian Personalized Ranking (BPR) loss~\cite{rendle2009bpr} to optimize the model parameters:
  \begin{equation}
\mathcal{L}_{\text{BPR}}(\mathbf{\Theta}) = \sum_{(u, i, j) \in \mathcal{D}} - \ln \sigma(\mathbf{e}_u^\top\mathbf{e}_i - \mathbf{e}_u^\top\mathbf{e}_j) ,
\label{eq2}
\end{equation}
where $\mathcal{D} = \{(u,i,j) | \mathbf{R}_{ui}=1,  \mathbf{R}_{uj}=0\}$ denotes the  training instances; $\sigma(\cdot)$ is the sigmoid function; $\mathbf{\Theta}$ is the model parameters. 

\begin{figure*}
\begin{center}\includegraphics[width=15.6cm]{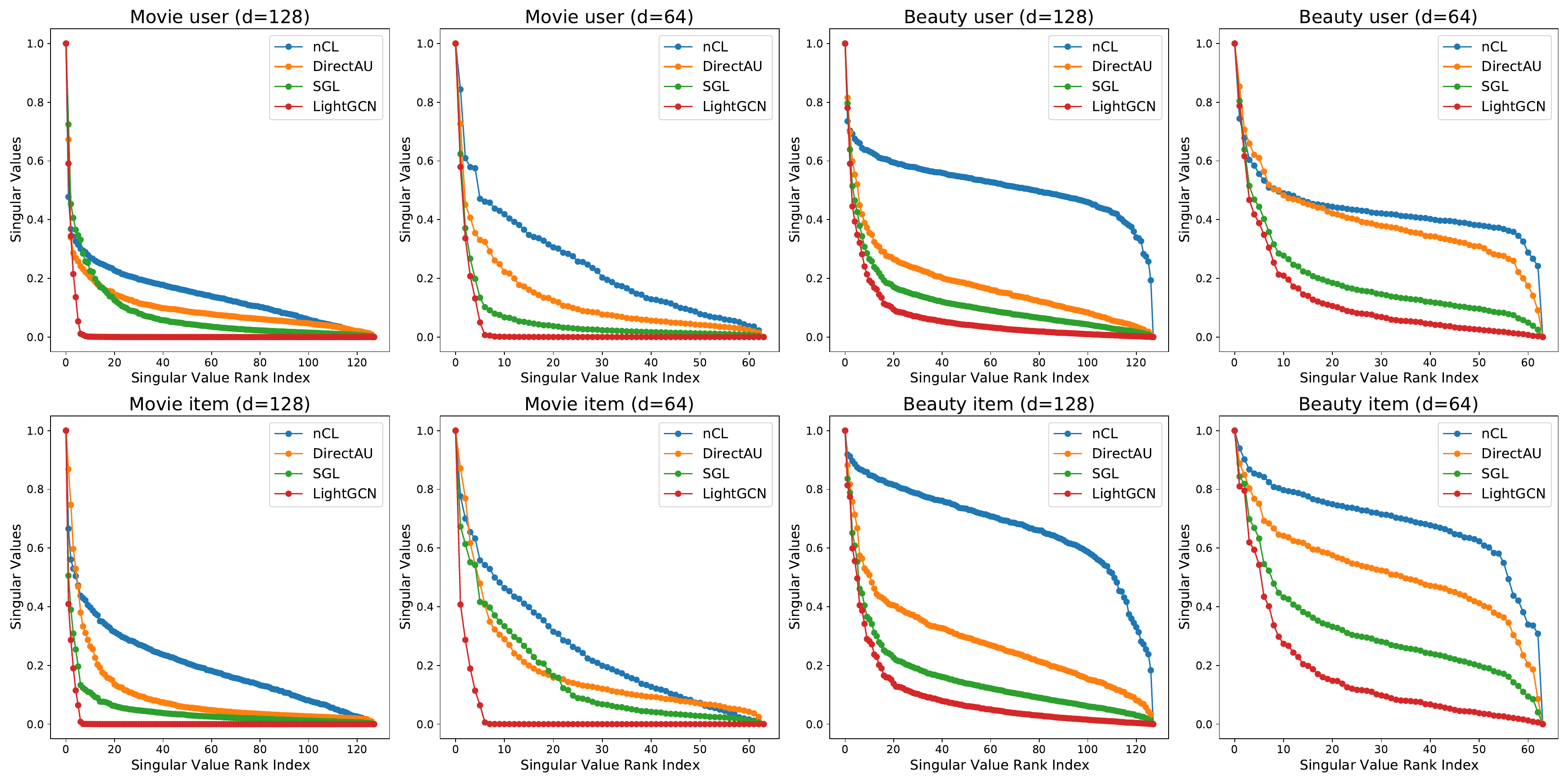}
	\end{center}
	\caption{Singular value spectrum of all the embeddings.    The spectrum contains the singular values of the covariance  matrix in  sorted  order. A number of singular values converge to zero (\textsl{e.g.}, LightGCN), indicating dimensional collapse.}
	\label{fig2}
 \vspace{-3mm}
\end{figure*}
\subsection{SGL}
SGL~\cite{wu2021self} extends LightGCN by  assuming that the high-quality user/item representations should be invariant to partial structure perturbations~\cite{chen2020simple,you2020graph}.  As such, SGL applies three augmentations: node dropout, edge dropout, and random walk, to create different views. Also, it treats the view of the same node as the positive pairs, \textsl{i.e.}, $\{(\mathbf{e}'_u, \mathbf{e}^{\prime\prime}_u) | u \in \mathcal{U}\}$, and the views of any different nodes as the negative pairs, \textsl{i.e.}, $\{(\mathbf{e}'_u, \mathbf{e}^{\prime\prime}_v) | u \in \mathcal{U}, u \not = v\}$.  SGL adopts  InfoNCE loss~\cite{oord2018representation} to maximize the agreement of  positive user pairs and disagreement of negative user pairs as:
  \begin{equation}
\mathcal{L}_{\text{ucl}}  = \sum_{u \in  \mathcal{U} } - \log \frac{\exp({\mathbf{e}^{\prime}_u}^\top\mathbf{e}^{\prime\prime}_u / \tau)}{\sum_{v \in \mathcal{U}}\exp({\mathbf{e}^{\prime}_u}^\top\mathbf{e}^{\prime\prime}_v / \tau)},
\label{eq3}
\end{equation}
where $\tau >0$ is the temperature. Analogously, one can obtain the contrastive loss for items $\mathcal{L}_{\text{icl}}$. Then, SGL  jointly optimizes Eq. (\ref{eq2}) and Eq. (\ref{eq3}) as:
$
\mathcal{L}_{\text{SGL}}  = \mathcal{L}_{\text{BPR}} + \lambda \cdot (\mathcal{L}_{\text{ucl}} + \mathcal{L}_{\text{icl}})$,
where $\lambda >0$   controls the strength between two loss functions.


\subsection{DirectAU}
Recently, several studies~\cite{wang2020understanding, gao2021simcse} identify two key properties highly related to the quality of CL-based representations: alignment and uniformity.  To achieve the two properties, DirectAU~\cite{wang2022towards} designs two learning loss functions as:
\begin{equation}
 \begin{gathered}
    \mathcal{L}_{\rm align} =\mathop{\mathbb{E}}_{(u, i)\sim p_{\rm pos}}\| \mathbf{e}_u- \mathbf{e}_i\|^2_F, \\
    \mathcal{L}_{\rm uniform} =\log\mathop{\mathbb{E}}_{(u,u')\sim p_{\rm user}}e^{-2\|\mathbf{e}_u - \mathbf{e}_{u'}
        \|^2_F} + \log\mathop{\mathbb{E}}_{(i,i')\sim p_{\rm item}}e^{-2\|\mathbf{e}_i  - \mathbf{e}_{i'} \|^2_F},
    \label{eq7}
\end{gathered}
\end{equation}
where $p_{\rm pos}$ denotes the distribution of positive user-item pairs; $ p_{\rm user}$ and $ p_{\rm item}$ are the distributions of users and items, respectively. Intuitively, $  \mathcal{L}_{\rm align}$ calculates expected distance between embeddings of the positive user-item pairs, while $ \mathcal{L}_{\rm uniform}$ measures how well the user/item embeddings are uniformly distributed.  To this end, DirectAU  straightly unifies the loss as: $\mathcal{L}_{\rm DirectAU} = \mathcal{L}_{\rm align} + \lambda \mathcal{L}_{\rm uniform}$, where $\lambda$ is a regularized parameter.

\subsection{Dimensional Collapse Problem}
In this section, we investigate the problem of dimensional collapse with different baselines. We choose LightGCN as the backbone for all baselines, with fixed number of layers ($L=2$), and vary the size of embeddings $d$ within $\{128, 64\}$ for users and items.  

We collect the well trained representations of users for all methods (all embedding vectors are normalized on the hypersphere), and simply  compute the users'   covariance matrix $\mathbf{C}_u \in \mathbb{R}^{d \times d}$ as:
  \begin{equation*}
\mathbf{C}_u = \frac{1}{|\mathcal{U}|} \sum_{u=1}^{|\mathcal{U}|} (\mathbf{e}_u -  \overline{\mathbf{e}}_u )(\mathbf{e}_u -  \overline{\mathbf{e}}_u )^\top,
\label{eq8}
\end{equation*}
where $\overline{\mathbf{e}}_u = \sum_{u=1}^{|\mathcal{U}|}\mathbf{e}_u$.  We calculate all singular values of $\mathbf{C}_u$ (\textsl{e.g.}, SVD($\cdot$)) in sorted order for all baselines, including the proposed nCL (details see Section 4). For easy comparison, all singular values are linearly  scaled into $[0,1]$  since we are more interested in the distribution (spectrum) to exploit the dimensional collapse.   The same process is applied to obtain the spectrum of items' singular values. Figure~\ref{fig2} shows the distributions of users' and items' singular values for  both MovieLens10M and Amazon Beauty datasets. 

In addition,   we utilize three metrics to evaluate various baselines: alignment, uniformity, and Recall@10. Alignment and uniformity are widely used metrics to assess the quality of embeddings learned in contrastive models, where lower values indicate better performance~\cite{wang2020understanding, gao2021simcse,wang2022towards}. On the other hand, Recall@10 is a metric that reflects the system's performance in top-$n$ recommendation, where higher values indicate better performance. Table \ref{t1} summarizes the results of these metrics. For each method, we report only the alignment and uniformity of user embeddings, as the results for item embeddings are similar and omitted in the table. From both Figure~\ref{fig2} and Table \ref{t1}, we have  following observations:
\begin{itemize}[leftmargin=*]
    \item We clearly find that all methods suffer from different levels of dimensional collapse,  especially for   LightGCN with the BPR loss. For example, only the first $10$ out of $128$ singular values have non-zero values, while others exactly converge to zeros. This indicates high redundancy and less information encoded by the learned dimension. Interestingly, simply reducing the size of dimension (\textsl{e.g.}, $128 \to 64$) does not solve the dimensional collapse problem.
    
    \item When comparing the spectrums of LightGCN and SGL, additional user and item InfoNCE losses are able to  prevent dimensional collapse but require complex data augmentation. Moreover,  DirectAU   optimizes both   alignment and uniformity. This mechanism  implicitly expands more volume of embedding space. 
    
    \item Among all the baselines, our nCL is the most effective approach to address the problem of dimensional collapse. This is because nCL enlarges the volume of embedding space by optimizing the rate-distortion function (Sec 4.2), which leads to the learning of diverse and discriminative user/item representations.
    \item Table \ref{t1} shows the lower alignment scores that demonstrate the smaller expected distances between the embeddings of user-item pairs, implying better performance.  Nevertheless, the lowest uniformity scores do not guarantee the best performance by comparing DirectAU and nCL. The uniformity enforces random instances should uniformly scatter on the hypersphere,  which may oversimplify the manifold of embedding spaces and have a worse dimensional collapse issue compared to our proposed nCL.
\end{itemize}

\begin{table}[t]
\small
  \centering
  \caption{Performance comparison of different methods with three metrics: alignment, uniformity, and Recall@10. For both alignment and uniformity, lower values are better, while higher  Recall@10 value indicates better performance.}
  \label{t1}
 \scalebox{0.95}{ \begin{tabular}{cccccc}
    \toprule
    \multicolumn{2}{c}{Setting} & \multicolumn{2}{c}{Beauty} & \multicolumn{2}{c}{MovieLen10M} \\
    \cmidrule(lr){1-2}\cmidrule(lr){3-4}\cmidrule(lr){5-6}
    Model & \multicolumn{1}{c}{Metric} & d=128 & d=64 & d=128 & d=64 \\
    \midrule
    \multirow{3}{*}{{LightGCN}} 
   & Alignment & 0.602 & 0.614 & 0.711 & 0.724 \\
   & Uniformity & -3.600& -3.524 &-1.159 &-1.135\\
   & Recall@10 & 0.0861 & 0.0863 & 0.1283 &  0.1306 \\
    \midrule
    \multirow{3}{*}{ {SGL}} 
   & Alignment & 0.596 & 0.602 & 0.667 & 0.701 \\
   & Uniformity & -3.624 & -3.601 & -2.887 & -2.654  \\
   & Recall@10 & 0.0900 & 0.0903 & 0.1730 & 0.1732 \\
    \midrule
    \multirow{3}{*}{ {DirectAU}} 
   & Alignment & 0.583 & 0.589 & 0.644 & 0.656 \\
   & Uniformity & \textbf{-3.912} & \textbf{-3.852} &\textbf{-3.754} &\textbf{-3.696} \\
   & Recall@10 & 0.1039 & 0.1041 & 0.2023 & 0.2026 \\
       \midrule
    \multirow{3}{*}{ {nCL}} 
   & Alignment & \textbf{0.578} & \textbf{0.586} & \textbf{0.637} & \textbf{0.644} \\
   & Uniformity & -3.887 & -3.724 & -3.601 &-3.589 \\
   & Recall@10 & \textbf{0.1164} & \textbf{0.1159} & \textbf{0.2314} & \textbf{0.2209} \\
    \bottomrule
  \end{tabular}}
  \vspace{-3mm}
\end{table}

In a nutshell,  all alignment, uniformity, and making full use of dimensional space are essentially  critical to obtain high performance according to aforementioned  analysis. More importantly, the experimental results also  suggest that sometimes we may  require a balanced trade-off among three properties in practice. Among the three properties, alignment and avoiding dimensional collapse are directly related to the system performance.

\section{The Proposed nCL}

Here we put forward  a new non-contrastive learning objective, named nCL,  that achieves  well alignment between positive user-item pairs and avoids dimensional collapse of user/item embedding spaces, without significantly sacrificing the uniformity of embeddings, leading to better performance for recommendation.

\begin{figure}[t!]
\centering
\includegraphics[width=5.0cm]{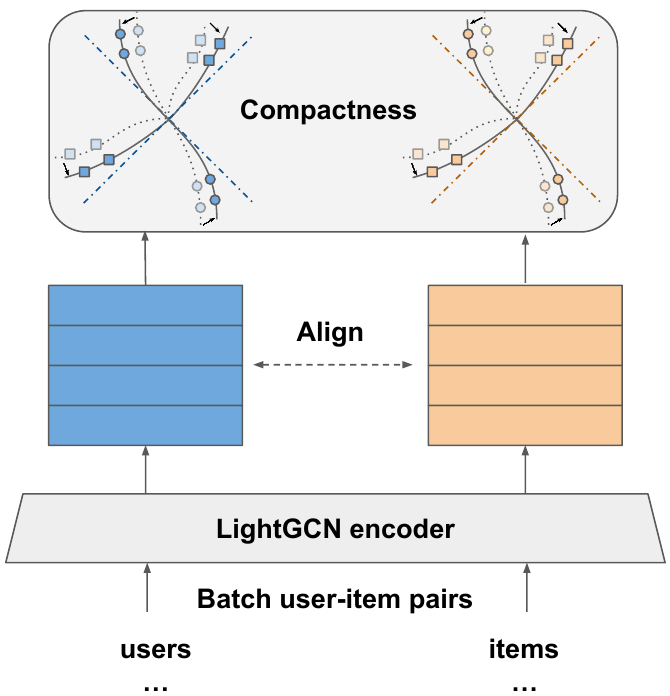}
\caption{Overview of the proposed nCL. nCL achieves two purposes: 1) alignment that minimizes distances between positive user-item pairs, and 2)  compactness  that encourages the  embeddings of users/items to expand  as much volume as possible, while simultaneously pushing each cluster (\textsl{e.g.}, squares and circles) to occupy as little space as possible. }
 \vspace{-3mm}
\label{fig3}
\end{figure}

Figure \ref{fig3} illustrates the overview of the proposed framework. our nCL adopts the LightGCN as backbone to generate the user/item embeddings: $\{\mathbf{e}_u \in \mathbb{R}^d|u\in \mathcal{U}\}$, and  $\{\mathbf{e}_i \in \mathbb{R}^d|i\in \mathcal{I}\}$. Then it explicitly optimizes two properties:  1) \textsl{Alignment} that minimizes distances between positive user-item pairs, and 2)  \textsl{Compactness}  that encourages the  embeddings of users/items to expand  as much volume as possible, while simultaneously pushing each cluster (\textsl{e.g.}, squares and circles) to occupy as little space as possible.

\subsection{\textbf{Alignment}}
The alignment loss tends to minimize the distances between
positive user-item pairs. To achieve this goal, we directly employ the distance between $\mathbf{e}_u$ and $\mathbf{e}_i$:  
\begin{equation}
    \mathcal{L}_{\rm align} =\mathop{\mathbb{E}}_{(u, i)\sim p_{\rm pos}}\| \mathbf{e}_u- \mathbf{e}_i\|^2_F
    \label{sds}
\end{equation}
where $p_{\rm pos}$ denotes the observed user-item interactions.  Despite its simplicity, the alignment loss is known to be surprisingly effective for many applications, such as the image-image alignments~\cite{wang2020understanding}, sentence-sentence alignments~\cite{gao2021simcse}, and user-item alignments~\cite{wang2022towards}.

\subsection{\textbf{ Compactness}}

In information theory, one can measure the \textsl{compactness} of representations via the rate-distortion function~\cite{thomas2006elements,han2022geometric}. Basically, the rate-distortion function $\mathcal{R}(\mathbf{z}, \epsilon)$  represents the minimum number of bits that is required to compress a random variable $\mathbf{z}$, such that the decompressing error is upper bounded by $\epsilon$. This function has been successful and is used to learn geometrically meaningful representations in computer vision recently~\cite{chan2022redunet,yu2020learning,zhang2021universal}.

Inspired by above findings, given  user representations $\mathbf{E}_u=[ \mathbf{e}_1, \mathbf{e}_2, \cdots, \mathbf{e}_{|\mathcal{U}|}] \in \mathbb{R}^{d \times |\mathcal{U}| }$, we can measure the compactness of user embeddings $\mathcal{R}(\mathbf{E}_u, \epsilon)$ as:
  \begin{equation}
\mathcal{R}(\mathbf{E}_u, \epsilon) = \frac{1}{2} \text{logdet}\left(\mathbf{I} + \frac{d}{|\mathcal{U}|\epsilon^2} \mathbf{E}_u \mathbf{E}_u^\top \right),
\label{eq9}
\end{equation}
where $\text{logdet}(\cdot)$ is log-determinant function, and  $\mathbf{I}$ is the  identity matrix that makes entire matrix $(\mathbf{I} + \frac{d}{|\mathcal{U}|\epsilon^2} \mathbf{E}_u \mathbf{E}_u^\top)$  positive definite. We next show how to optimize $\mathcal{R}(\mathbf{E}_u, \epsilon)$  to avoid dimensional collapse.

Using  SVD, we have $\mathbf{E}_u = \mathbf{U}\mathbf{S}\mathbf{V}^\top$, where $\mathbf{S} = \text{diag} (s_1, s_2 , \dots, s_d)$ contains the singular values, let $t= \frac{d}{|\mathcal{U}|\epsilon^2}$, then
 \begin{equation*}
  \begin{aligned}
 \text{logdet}(\mathbf{I} + t\mathbf{E}_u \mathbf{E}_u^\top ) &=  \text{logdet}(\mathbf{U}(\mathbf{I} + t \mathbf{S}^2)\mathbf{U}^\top ) = \text{logdet}(\mathbf{I} + t \mathbf{S}^2 )  \\
 & = \text{log}\prod_{i=1}^{d} (1 + t s^2_i) = \sum_{i=1}^{d} \text{log} (1 + t s^2_i).
  \end{aligned}
\label{eq10}
\end{equation*}

Above equality reveals that maximizing $\mathcal{R}(\mathbf{E}_u, \epsilon)$ prevents the singular values of $\mathbf{E}_u$ and its covariance matrix converging to zeros, avoiding dimensional collapse.  In other words, maximizing $\mathcal{R}(\mathbf{E}_u, \epsilon)$ encourages the  entire embeddings $\mathbf{E}_u$ to expand  as much volume as possible. This can be verified by the singular spectrum of our proposed nCL in Figure~\ref{fig2}.

In practice, the vectors $\mathbf{E}_u$ can be from a mixture distribution, \textsl{e.g.}, the users can be clustered into several groups. As suggested by recent studies~\cite{chan2022redunet,yu2020learning}, we can measure the rate distortion for each cluster. In particular, we assume the user embeddings $\mathbf{E}_u$ can be partitioned into $K$ groups, with the probability matrix $\bm{\pi} \in \mathbb{R}^{|\mathcal{U}|\times K}$, where ${\pi}_{uk} \in [0, 1]$ implies the probability of user $u$ assigned to cluster $k$, and $\sum_{k=1}^{K}{\pi}_{uk} =1$.  
We define the membership matrix for cluster $k$ as: $\bm{\Pi}_k = \text{diag}(\pi_{1k},\cdots, \pi_{|\mathcal{U}|k} ) \in \mathbb{R} ^{|\mathcal{U}| \times |\mathcal{U}|}$, and summarize group information into $\bm{\Pi} = \{\bm{\Pi}_k\}_{k=1}^K$. Then, the average per-cluster coding rate can be written as:
  \begin{equation}
\mathcal{R}^c(\mathbf{E}_u, \epsilon|\bm{\Pi}) = \sum_{k=1}^K\frac{\Tr(\bm{\Pi}_k)}{2|\mathcal{U}|} \cdot \text{logdet}\left(\mathbf{I} + \frac{d}{\Tr(\bm{\Pi}_k)\epsilon^2} \mathbf{E}_u \bm{\Pi}_k \mathbf{E}_u^\top \right),
\label{eq11}
\end{equation}
where $\Tr(\cdot)$ is the trace operation. As the embeddings in same cluster should be coherent, we can minimize the term $\mathcal{R}^c(\mathbf{E}_u, \epsilon|\bm{\Pi})$. By doing so, it would push embeddings in each cluster to occupy as little space as possible. To this end, we can optimize the compactness of user embeddings $\mathbf{E}_u$ by minimizing:
  \begin{equation}
\mathcal{L}_\text{compact}(\mathbf{E}_u) = \mathcal{R}^c(\mathbf{E}_u, \epsilon|\bm{\Pi}) - \mathcal{R}(\mathbf{E}_u, \epsilon).
\label{eq12}
\end{equation}
Intuitively, minimizing the first term  seeks to 
compress the volume of the embeddings for each cluster, while the second term aims to  expand the overall volume of the embeddings. As such, it encourages more discriminative representations across different clusters, simultaneously, more compact representations within same clusters. For items, $\mathcal{L}_\text{compact}(\mathbf{E}_i)$ can be defined in a similar fashion.

The rest question becomes how to effectively design the membership matrices $\bm{\Pi}$ for Eq. (\ref{eq12}). We next propose two simple strategies to obtain the cluster information.

\subsection{\textbf{Cluster Information}}
 \subsubsection*{\textbf{Option I (nCLG)}}
 Real-world datasets rarely  provide the cluster information for users/items. Here we propose to create the membership matrices using  co-occurrence graph topology, named nCLG. For users, we can first obtain the user-user graph from the user-item interaction matrix: $\mathbf{A} = \mathbf{R} \mathbf{R}^\top \in \mathbb{R}^{|\mathcal{U}| \times |\mathcal{U}| }\}$, where $\mathbf{A}_{uv}$ denotes the proximity between  $u$ and $v$. Normally, a large proximity score indicates that users $u$ and $v$ come from same cluster with high probability. Such co-occurrence graph has been shown to be useful for  recommender systems~\cite{mao2021ultragcn}.
 
Contrastive learning treats two samples as different classes as long as they are from different instances. With this analogy in mind, we assume that users $u$ and $v$ are from different clusters if $\mathbf{A}_{uv} < \eta$, otherwise they are from the same cluster, where $\eta$ is a threshold. Therefore, we can define the membership matrix $\bm{\Pi}_k = \text{diag}(\mathbb{I}(\mathbf{a}_k \ge \eta)) \in \mathbb{R}^{|\mathcal{U}| \times |\mathcal{U}| }$, where $\mathbf{a}_k$ is the $k$-th column of $\mathbf{A}$ and $\mathbb{I}(\cdot)$ denotes the indicator function. The collection $\bm{\Pi} = \{\bm{\Pi}_k | 1\le k \le |\mathcal{U}|\}$   can be then used in  Eq. (\ref{eq12}).

For membership matrices $\bm{\Pi}  $, we need to compute per-cluster coding rate $|\mathcal{U}|$ times, which is not scalable for large dataset. Instead of using all $\bm{\Pi}_k$, we randomly sample fixed number of $\bm{\Pi}_k$, \textsl{e.g.,} $50$,  during each batch training. The similar algorithm can be applied to obtain the membership matrices for items.

 \subsubsection*{\textbf{Option II (nCL)}} nCLG only considers the first-order node-node proximity, a more sophisticated approach is to learn the semantic structure of users/items via deep clustering~\cite{caron2020unsupervised,asano2019self,wang2022clusterscl}. Suppose the user embeddings $\mathbf{E}_u=[ \mathbf{e}_1, \mathbf{e}_2, \cdots, \mathbf{e}_{|\mathcal{U}|}] \in \mathbb{R}^{d \times |\mathcal{U}| }$, drawn from a space of $K$ possible labels (clusters), we can apply a  multilayer perceptron network  $\Phi: \mathbb{R}^d \to \mathbb{R}^K$ to compute the labels $y_i, y_2, \cdots, y_{|\mathcal{U}|} \in  \{1, \cdots, K\}$, via the softmax : $p(y=\cdot|\mathbf{e}_u) = \text{softmax}(\Phi(\mathbf{e}_u))$. As a result, the model parameters can be learned by minimizing the average cross-entropy loss:
   \begin{equation*}
E(p|y_i, y_2, \cdots, y_{|\mathcal{U}|}) = -\frac{1}{|\mathcal{U}|}\sum_{u=1}^{|\mathcal{U}|}\log p(y_u|\mathbf{e}_u).
\label{eq13}
\end{equation*}
 Learning above objective, however, requires a labelled dataset. Inspired by the self-labelling mechanism~\cite{caron2020unsupervised}, we can define the labels as posterior distribution $q(y|\mathbf{e}_u)$, and rewrite above equation as:
    \begin{equation}
E(p,q) = -\frac{1}{|\mathcal{U}|}\sum_{u=1}^{|\mathcal{U}|} \sum_{y=1}^{K}q(y|\mathbf{e}_u)\log p(y|\mathbf{e}_u).
\label{eq14}
\end{equation}
 Without any constraints, Eq. (\ref{eq14}) will lead to a degenerate solution, \textsl{i.e.}, all samples are assigned to one cluster with same label. To address this problem, let $\mathbf{P}_{yu} = p(y|\mathbf{e}_u)\frac{1}{|\mathcal{U}|}$ and $\mathbf{Q}_{yu} = q(y|\mathbf{e}_u)\frac{1}{|\mathcal{U}|}$ be two $K \times |\mathcal{U}|$ matrices, we enforce an equal partition of clusters by constraining $\mathbf{Q}$ to belong to the \textsl{transportation polytope}~\cite{cuturi2013sinkhorn}. As such, Eq. (\ref{eq14}) becomes:
 \begin{equation}
 \begin{aligned}
 \min_{\mathbf{Q}}   \langle  \mathbf{Q}, &-\log\mathbf{P}  \rangle, \\
 \text{s.t.}  \qquad   \mathbf{Q} \bm{1}_{|\mathcal{U}|}   = \frac{1}{K} \bm{1}_{K}, & \quad  \mathbf{Q}^\top \bm{1}_{K}   = \frac{1}{|\mathcal{U}|} \bm{1}_{|\mathcal{U}|},
 \end{aligned}
\label{eq15}
\end{equation}
 where $\bm{1}_{|\mathcal{U}|}$ is the vector of ones in dimension $|\mathcal{U}|$.  Once $\mathbf{Q}^*$ is found, we can regard it as the cluster assignments, and compute the membership matrices $\bm{\Pi} = \{\bm{\Pi}_k\}_{k=1}^K$ as shown in Sec 3.2.
 
 In this work, we choose the recent Inexact Proximal point method for Optimal Transport (IPOT) algorithm~\cite{xie2020fast} to compute the optimal matrix $\mathbf{Q}^*$ in Eq. (\ref{eq15}). Unlike the Sinkhorn solver~\cite{cuturi2013sinkhorn}, it does not need to back propagate the gradients through the proximal point iterations due the the Envelope Theorem~\cite{xie2020fast}. This significantly accelerates the learning process and improves training stability.

	\begin{algorithm}
 \small
		\DontPrintSemicolon 
		\KwIn{ user-item interactions data $\mathbf{R}$;\\ 
      regularized parameter $\alpha$; \\the number of latent clusters $K$; \\$\epsilon$ for rate-distortion function;
      embedding dimension $d$.}
		Initialize model parameters $\mathbf{\Theta}$;\;
		\For{\text{each epoch}}{
			\For{each mini-batch  pairs $\{(u,i)\}\in\mathcal{R}$}{
				Compute the embeddings $\mathbf{E}_u$ and $\mathbf{E}_i$ via LightGCN;\;
				Normalize $\mathbf{E}_u$ and $\mathbf{E}_i$ into unit vectors; \;
				Compute the loss $\mathcal{L}_{\text{nCL}}$ in Eq. (\ref{eq16});\;
			Update the model parameters $\mathbf{\Theta}$ via SGD;
				
			}
			Update  membership matrices  $\bm{\Pi} = \{\bm{\Pi}_k\}_{k=1}^K$ via IOPT solver;\;
		}
		\caption{nCL}
		\KwOut{ Well-trained model parameters $\mathbf{\Theta}$.}
	\end{algorithm}
 
\subsection{\textbf{Overall Loss}}
To this end, combining the alignment $\mathcal{L}_{\rm align}$ Eq. (\ref{sds}) and the compactness $\mathcal{L}_\text{compact}$  Eq. (\ref{eq12}) yields our objective function as follows:
  \begin{equation}
\mathcal{L}_{\text{nCL}}  = \mathcal{L}_{\text{align}} + \alpha \cdot (\mathcal{L}_\text{compact}(\mathbf{E}_u) + \mathcal{L}_\text{compact}(\mathbf{E}_i)),
\label{eq16}
\end{equation}
where $\alpha>0$   controls the balance between two alignment and compactness. Noted that we separately measure the compactness of user representations
and item representations because the data distribution of users and
items might be diverse. Algorithm 1 summarizes the learning steps of nCL.

Finally, it is worth mentioning that our nCL framework is non-contrastive, which eliminates the need for complex data augmentations and enables small batch training. These appealing properties make it easier to implement compared to contrastive frameworks.

\subsubsection*{\textbf{Model Complexity}}

The  main complexity of nCL comes from the computation of  $\text{logdet}(\mathbf{I} + \frac{d}{|\mathcal{U}|\epsilon^2} \mathbf{E}_u \mathbf{E}_u^\top)$ in $\mathcal{L}_{\text{nCL}}$, which takes $\mathcal{O}(d^3 + {|\mathcal{U}|}d^2)$. However, the size of dimension $d$ is often very small $d\ll |\mathcal{U}|$. Our nCL thus has a linear complexity with respect to the number of users/items and has fast convergence rate (see Figure~\ref{ddds} ), leading to the scalability for large dataset. Also, we are aware that recent advanced algorithms are capable of speeding up the computation using variational function~\cite{baek2022efficient} and we plan to investigate this as part of future work.

\section{Experiments}

\begin{table}
\small
  \centering
  \caption{Statistics of four benchmark 
 datasets.}
  \label{tab:dataset}
  \begin{tabular}{lccccc}
    \toprule
    Dataset & \tabincell{c}{\#user\\($|\mathcal{U}|$)} & \tabincell{c}{\#item\\($|\mathcal{I}|$)} & \tabincell{c}{\#inter.\\($|\mathcal{R}|$)} & \tabincell{c}{avg. inter.\\per user} & density\\
    \midrule
    Movie & 72.0k & 10.0k & 10.0m & 138.9 & 1.39\%\\
    Beauty & 22.4k & 12.1k & 198.5k & 8.9 & 0.07\%\\
    Book & 52.6k & 91.6k & 2984.1k & 56.7 & 0.06\%\\
    Yelp & 31.7k & 38.0k & 1561.4k & 49.3 & 0.13\%\\
  \bottomrule
\end{tabular}
  \vspace{-3mm}
\end{table}

\subsection{Experimental Settings}
\subsubsection{\textbf{Datasets}}
We use four public benchmark datasets to evaluate
recommendation performance: \textbf{MovieLens10M}\footnote{https://grouplens.org/datasets/movielens/}, \textbf{Beauty},  \textbf{Book}, and \textbf{Yelp}\footnote{https://www.yelp.com/dataset}, where both Beauty and Book are from Amazon dataset\footnote{https://jmcauley.ucsd.edu/data/amazon/links.html}. Following previous work~\cite{he2020lightgcn, wang2022towards},  we adopt the $5$-core settings and 
the statistics of datasets  are summarized in Table \ref{tab:dataset}.

\begin{table*}
\tabcolsep=6.5pt
\centering
\caption{Top-\textit{n} performance on four datasets. The best results are in bold face with $*$, the second best ones are in bold face, and third best ones are underlined. The Improv 
 denotes the relative improvements of nCL over the best baselines (except nCLG).}
\scalebox{0.85}{\begin{tabular}{C{1cm}L{1.6cm}ccccccccccr}
    \toprule
    \multicolumn{2}{c}{Settings} & \multicolumn{8}{c}{Baselines} & \multicolumn{3}{c}{Ours}\cr 
    \cmidrule(lr){1-2} \cmidrule(lr){3-10} \cmidrule(lr){11-13}
    Dataset & \multicolumn{1}{c}{Metric} & BPRMF  & LightGCN & DGCF & BUIR & CLRec & SGL & SimGCL & DirectAU    & nCLG & nCL & Improv. \cr
    \midrule
    \multirow{6}{*}{\rotatebox{90}{MovieLens10M}} 
    & Recall@10 & 0.1264  &  0.1306 &0.1569  &0.1375 &0.1754  &0.1732   &0.1871     & \underline{0.2026} & \textbf{0.2295} & \textbf{0.2314}$^{*}$   &+14.21\%  \cr
    & Recall@20 & 0.1877 & 0.2053 &0.2435 &0.2210  &0.2589   &0.2437  &0.2642    & \underline{0.3028} & \textbf{0.3222} & \textbf{0.3301}$^{*}$ &+9.02\%\cr
    & Recall@50 & 0.2910  & 0.3284  &0.3640  &0.3307  &0.3610  &0.3561 &0.3784  & \underline{0.4522} & \textbf{0.4647} & \textbf{0.4752}$^{*}$  &+5.09\%\cr
    \cmidrule(lr){2-2} \cmidrule(lr){3-10} \cmidrule(lr){11-13}
    & NDCG@10 & 0.1543 & 0.1766 &0.1874 &0.1803 &0.1904     &0.1877  &0.1950     & \underline{0.2265} & \textbf{0.2582} & \textbf{0.2596}$^{*}$  &+14.61\%\cr
    & NDCG@20 & 0.1704  & 0.1892 &0.2018  &0.1924  &0.2078   &0.1991  &0.2239    & \underline{0.2488} & \textbf{0.2762} & \textbf{0.2814}$^{*}$ &+13.10\% \cr
    & NDCG@50 & 0.1986  & 0.2212 &0.2674  &0.2411  &0.2499    &0.2444   &0.2585 & \underline{0.2918} & \textbf{0.3200} & \textbf{0.3284}$^{*}$  &+12.54\%\cr
    \midrule
    \multirow{6}{*}{\rotatebox{90}{Beauty}} 
    & Recall@10 & 0.0806  & 0.0863 & 0.0897 & 0.0816 & {0.0937}  &0.0903    &0.0939    & \underline{0.1041} & \textbf{0.1116} & \textbf{0.1164}$^{*}$ &+11.82\% \cr
    & Recall@20 & 0.1153  & 0.1201 & 0.1283 & 0.1204 & {0.1337}  &0.1228   &0.1367    & \underline{0.1434 }& \textbf{0.1516} & \textbf{0.1570}$^{*}$  &+9.48\%\cr
    & Recall@50 & 0.1763  & 0.1819 & 0.1958 & 0.1866 & {0.1996}  &0.1964  &0.2028    & \underline{0.2048} & \textbf{0.2173} & \textbf{0.2232}$^{*}$  &+8.98\% \cr
    \cmidrule(lr){2-2} \cmidrule(lr){3-10} \cmidrule(lr){11-13}
    & NDCG@10 & 0.0444 & 0.0484 & 0.0501 & 0.0457 & {0.0547}    &0.0521    &0.0595    & \underline{0.0635} & \textbf{0.0650} & \textbf{0.0671}$^{*}$ &+5.67\% \cr
    & NDCG@20 & 0.0534  & 0.0581 & 0.0600 & 0.0556 & {0.0651}   &0.0644    &0.0682    & \underline{0.0740} & \textbf{0.0755} & \textbf{0.0763}$^{*}$  &+3.11\%\cr
    & NDCG@50 & 0.0658  & 0.0699 & 0.0738 & 0.0692 & {0.0786}   &0.0734    &0.0792    & \underline{0.0871} & \textbf{0.0890} & \textbf{0.0902}$^{*}$ &+3.56\% \cr
    \midrule
    \multirow{6}{*}{\rotatebox{90}{Book}} 
    & Recall@10  & 0.0574 & 0.0601 &0.0624 &0.0568 &0.0629  &0.0627  & \underline{0.0631}    & 0.0620 & \textbf{0.0728} & \textbf{0.0732}$^{*}$  &+16.01\%\cr
    & Recall@20  & 0.0802 & 0.0875 &0.0898 &0.0796 &0.0912   &0.0919  &0.0941    & \underline{0.0966} & \textbf{0.1090} &\textbf{0.1113}$^{*}$  &+15.22\%\cr
    & Recall@50  & 0.1031 &0.1289 &0.1382 &0.1027    &0.1479  &0.1477 &0.1533 & \underline{0.1659} & \textbf{0.1780} &\textbf{0.1802}$^{*}$  &+8.62\%\cr
    \cmidrule(lr){2-2} \cmidrule(lr){3-10} \cmidrule(lr){11-13}
    & NDCG@10  & 0.0450 & 0.0496 &0.0513 &0.0432 &0.0517     &0.0520  &\underline{0.0543}    & 0.0516 & \textbf{0.0610} & \textbf{0.0623}$^{*}$  &+14.73\%\cr
    & NDCG@20 & 0.0507 & 0.0576 &0.0588 &0.0497 &0.0597    &0.0606   &\underline{0.0642}      & 0.0629 & \textbf{0.0728} & \textbf{0.0741}$^{*}$  &+15.42\%\cr
    & NDCG@50  & 0.0716 &0.0752 &0.0797 &0.0711    &0.0808  &0.0802  &0.0815     & \underline{0.0831} & \textbf{0.0928} & \textbf{0.0942}$^{*}$  &+13.48\%\cr
    \midrule
    \multirow{6}{*}{\rotatebox{90}{Yelp}} 
    & Recall@10  & 0.0416 & 0.0508 & 0.0519 & 0.0444 & 0.0547   &0.0545    &0.0582          & \textbf{0.0676} & \underline{0.0641} & \textbf{0.0695}$^{*}$    &+2.81\%\cr
    & Recall@20  &  0.0693 & 0.0833 & 0.0849 & 0.0737 & 0.0890  &0.0896     &0.0937         & \textbf{0.1075} & \underline{0.1043} & \textbf{0.1120}$^{*}$   &+4.19\%\cr
    & Recall@50  & 0.1293  & 0.1534 & 0.1575 & 0.1386 & 0.1606  &0.0166     &0.1733        &  \textbf{0.1911} & 0.1855 & \textbf{0.1975}$^{*}$               &+3.35\%\cr
    \cmidrule(lr){2-2}\cmidrule(lr){3-10} \cmidrule(lr){11-13}
    & NDCG@10  & 0.0335 & 0.0406 & 0.0409 & 0.0349 & 0.0436   &0.0441      &0.0483        &  \textbf{0.0547} & \underline{0.0524} & \textbf{0.0564}$^{*}$    &+3.11\%\cr
    & NDCG@20 & 0.0428  & 0.0514 & 0.0521 & 0.0448 & 0.0551    &0.0554      &0.0571     & \textbf{0.0679} & \underline{0.0658} & \textbf{0.0698}$^{*}$       &+2.80\%\cr
    & NDCG@50 &0.0602 & 0.0717 & 0.0732 & 0.0636 & 0.0758    & 0.0782      &0.0814      & \textbf{0.0920} & \underline{0.0895} & \textbf{0.0965}$^{*}$     &+4.89\%\cr
    \bottomrule
\end{tabular}}
\label{tab4}
\end{table*}

\subsubsection{\textbf{Baselines}}
We compare the performance of our nCL with various state-of-the-art  methods:
\textbf{BPRMF}~\cite{rendle2009bpr}: A matrix factorization model with the Bayesian personalized ranking loss. \textbf{LightGCN}~\cite{he2020lightgcn}: A GNN model that performs linear propagation between neighbors on the user-item bipartite graph.
\textbf{DGCF}~\cite{wang2020disentangled}: A  disentangled GNN method that models the intent-aware interaction graphs. \textbf{BUIR}~\cite{lee2021bootstrapping}: A  negative-sample-free  method that learns user and item embeddings with only positive interactions. \textbf{CLRec}~\cite{zhou2021contrastive}: A contrastive model that adopts the InfoNCE loss to address the exposure bias in recommender systems. \textbf{SGL}~\cite{wu2021self}: A model that introduces additional InfoNCE loss to enhance recommendation. We select edge drop to perform augmentation due to its high performance. \textbf{SimGCL}~\cite{yu2022graph}: A GNN model that inserts random noises   as data augmentation with contrastive InfoNCE loss.  \textbf{DirectAU}~\cite{wang2022towards}: This is the state-of-the-art CF method that directly optimizes alignment and uniformity. \textbf{nCLG} and \textbf{nCL}: our proposed methods with different strategies for computing membership matrices.


\subsubsection{\textbf{Evaluation Protocols}}
Following the settings~\cite{wang2022towards}, for each dataset, we randomly split each user's historical interactions into training/validation/test sets with the ratio of 8:1:1. To evaluate the performance of top-\textit{n} recommendation, we use two common metrics: 
Recall@\textit{n} and NDCG@\textit{n}, where $n=[10,20,50]$ in the experiments.
To avoid sampling bias, we considered the ranking list of all items, as suggested by~\cite{DBLP:conf/kdd/KricheneR20}. We repeated each experiment 10 times independently and reported the average scores.

\subsubsection{\textbf{Implementation Details}}
We implement all the methods using the PyTorch, and Adam  is used as the default optimizer. Early stop is adopted if NDCG@10 on the validation dataset continues to drop for 10 epochs.
We search the embedding size within $\{64, 128, 256\}$, and the  learning rate is set to 1e$^{-3}$ for all the methods. The training batch size is searched within $\{512, 1024, 2048, 4096\}$.  For all baselines, their hyperparameters are initialized as in their original papers and are then carefully tuned to achieve optimal performance.  For our nCL, we tune  $\epsilon^2$ for rate-distortion function within $\{0.01, 0.05, 0.1, 0.5\}$, the number of user/item clusters $K$ within $\{100, 200, 300, 400, 500\}$, the regularized parameter $\alpha$ in Eq. (\ref{eq16}) within $\{0.01, 0.05, 0.1, 0.5, 1\}$, in the experiments.

\subsection{Overall Performance}
 Table~\ref{tab4} shows the performance comparison of the proposed nCL and other baselines on four public datasets. Overall, our proposed nCL consistently yields the best performance across all cases. Additionally, we have the following observations:
\begin{itemize}[leftmargin=*]
    \item Comparing to the traditional CF methods BPRMF and LightGCN, contrastive methods (e.g., CLRec, SGL, SimGCL) generally obtain better results. For example, with additional InfoNCE loss, SGL and SimGCL outperform their backbone LightGCN, indicating the benefits of  self-supervised signals in recommendation.
    
    \item Different  data augmentation techniques have different impacts on the final results. For example, SimGCL performs better than SGL, but worse than DirectAU and nCL in many cases. This implies that a good design of data augmentation is very important in contrastive learning. Nevertheless, most of the existing data augmentation methods  are heuristic  without any theoretical analysis (\textsl{e.g.}, dropping randomly node/edge in SGL, adding uniform noise in SimGCL), which suffers from more uncertainty compared to non-contrastive methods.
    \item  Both DirectAU and nCL are non-contrastive models, which however have better performance than others like SGL, CLRec, and SimGCL. This suggests that designing a high-quality loss function is able to achieve a similar purpose of contrastive models, like considering alignment, uniformity, and avoiding dimensional collapse. Also, nCL achieves the best performance, owing to its ability to prevent dimensional collapse.
\end{itemize}

\begin{figure}[t!]
\centering
\includegraphics[width=7.4cm]{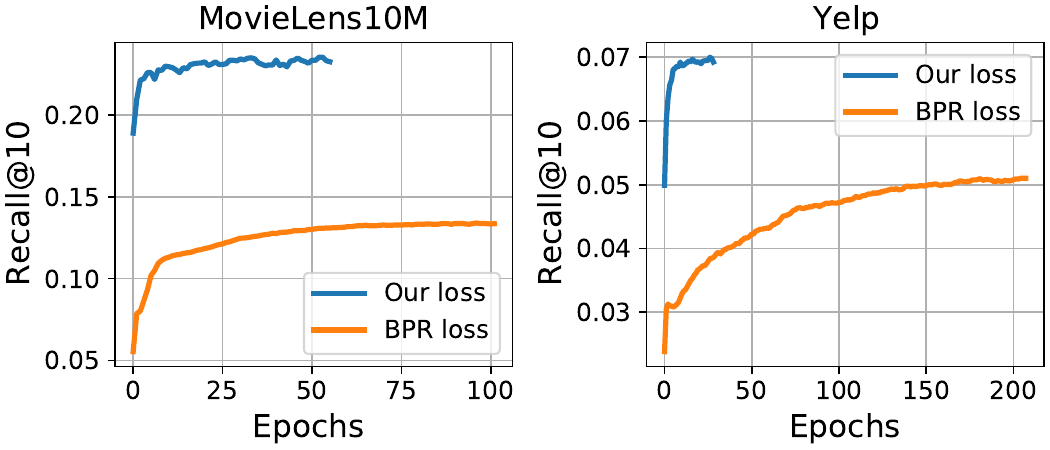}
\caption{ The training curves (testing recall@10) of our loss and the BPR loss for MovieLens10M and Yelp datasets. 
 }
\label{ddds}
\vspace{-3.6mm}
\end{figure}

\noindent \textbf{Training Efficiency.} We further plot the training curves of LightGCN and our nCL with early stop training strategy.   The  primary difference between these two is the
 loss function (BPR loss vs. our loss in Eq.(\ref{eq16})).  From Figure~\ref{ddds}, we notice that our loss function is able to accelerate the training of LightGCN significantly. Taking Yelp dataset as the example, our loss function only takes around 30 epochs to converge, while the BPR loss takes around 220 epochs. In terms of running time,  the time taken for training per epoch of LightGCN and nCL are about 40s and 74s,  respectively. Considering both convergence speed and running time per epoch, our proposed nCL speeds up the LightGCN by a factor of $\times 3.96$, indicating the superiority of our method in GNNs training acceleration.

\subsection{Study of nCL}
\subsubsection{\textbf{Sparse Recommendation}}
\begin{figure}
\centering
\includegraphics[width=7.4cm]{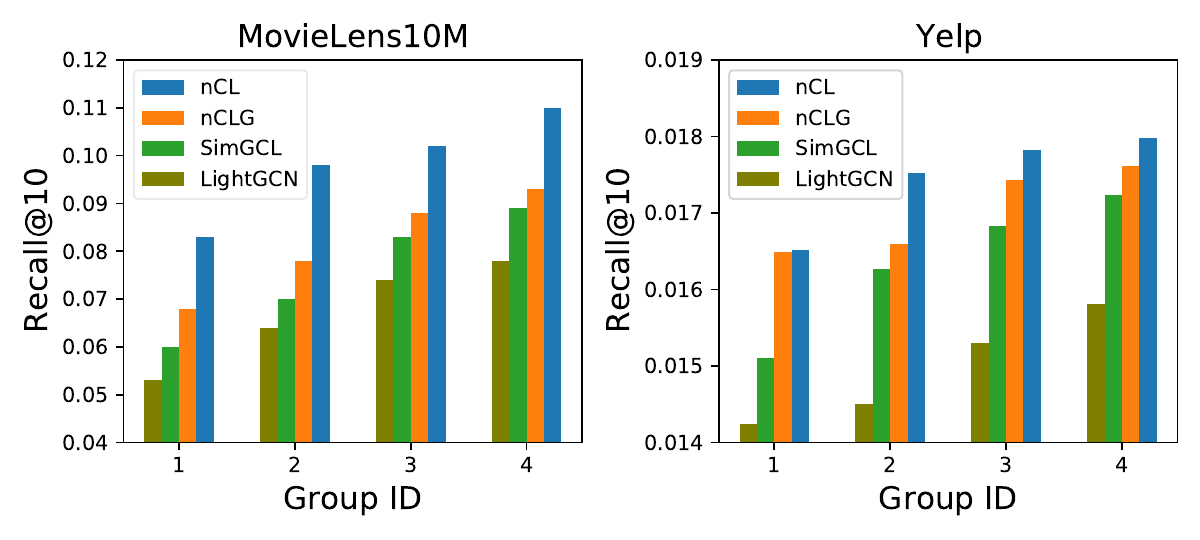}
\caption{ Performance analysis (Recall@10) for different levels of sparsity  for MovieLens10M and Yelp datasets.
 }
\label{s}
\end{figure}

As graph convolutions are essentially local operators, they
 have limited benefits for long-tailed nodes  to collect enough information.  To  verify  whether the proposed nCL is  able to  alleviate the data sparsity issue, we sample items into four groups based on popularity, while keeping the total number of interactions of each group the same~\cite{yu2022graph}. The larger the GroupID is, the larger degrees the items have.  For example, Group 1 contains the items with degree $[0, 100)$, Group 2 with  $[100, 200)$, Group 3 with  $[200, 300)$ and Group 4 with  $[300, 400)$. 

We perform experiments on both the MovieLens10M and Yelp datasets, with Recall@10 as the evaluation metric. From the results presented in Figure~\ref{s}, we can observe that the performance of nCL and nCLG consistently outperforms LightGCN and SimGCL for different levels of sparsity. This indicates that our algorithms, which consider the alignment and compactness of the embeddings, have great potential for long-tail recommendation, even without any data augmentations as in CL-based methods.

\subsubsection{\textbf{Ablation Study}}  Our objective function $\mathcal{L}_{\text{nCL}}$ Eq. (\ref{eq16}) contains two parts: alignment $\mathcal{L}_{\text{align}}$  and compactness $\mathcal{L}_{\text{compact}}$. The $\mathcal{L}_{\text{align}}$ connects the embedding spaces of users and items, which should be included in order to predict the links between two types of nodes. Here we mainly exploit the impacts of $\mathcal{R}^c(\cdot)$ and $\mathcal{R}(\cdot)$ in the $\mathcal{L}_\text{compact}(\cdot)$~Eq. (\ref{eq12}) via an ablation study. Table~\ref{tt4} shows the performance of our default method and its four variants. 

Not surprisingly, removing any component of $\mathcal{L}_\text{compact}(\cdot)$ impairs the performance. Also, we notice that removing $\mathcal{R}(\mathbf{E}_u, \epsilon)$ or $\mathcal{R}(\mathbf{E}_i, \epsilon)$  significantly hurts the performance. Presumably this is because the $\mathcal{R}(\mathbf{E}_u, \epsilon)$ or $\mathcal{R}(\mathbf{E}_i, \epsilon)$ encourages the entire embeddings to expand as much volume as possible, which prevents dimensional collapse. Meanwhile, the impacts of $ \mathcal{R}^c(\mathbf{E}_u, \epsilon|\bm{\Pi}_u)$ and $ \mathcal{R}^c(\mathbf{E}_i, \epsilon|\bm{\Pi}_i)$  can not be ignored since it addresses the community embedding spaces of users/items. We believe that, with auxiliary information like item categories information, we  can further boost the performance using more accurate membership matrix $\bm{\Pi}_u$/$\bm{\Pi}_i$.

\begin{table}[]
\small
\caption{Ablation analysis of  the compactness of representations in nCL, where $\downarrow$ indicates a severe performance drop.}
\begin{tabular}{l|cc|cc}
\midrule
     Architecture   & \multicolumn{2}{c|}{MovieLens10M}                            & \multicolumn{2}{c}{Yelp}                                    \\
        & \multicolumn{1}{c}{R@10} & \multicolumn{1}{c|}{N@10} & \multicolumn{1}{c}{R@10} & \multicolumn{1}{c}{N@10} \\ \midrule
(0) Default nCL &           0.2314                    &      0.2596                       &     0.0695                          &          0.0564                   \\
(1) Remove $\mathcal{R}(\mathbf{E}_u, \epsilon)$  &    0.1896$\downarrow$                          &   0.2113$\downarrow$                        &          0.0574$\downarrow$                     &          0.0501                   \\
(2) Remove $ \mathcal{R}^c(\mathbf{E}_u, \epsilon|\bm{\Pi}_u)$ &     0.1932                          &      0.2174                       &      0.0587                         &              0.0490               \\
(3) Remove $\mathcal{R}(\mathbf{E}_i, \epsilon)$ &        0,1954                       &          0.2242                   &             0.0580                  &                   0.0481          \\
(4) Remove $ \mathcal{R}^c(\mathbf{E}_i, \epsilon|\bm{\Pi}_i)$ &       0.2033                        &         0.2301                    &            0.0593                   &              0.0472$\downarrow$           \\ \midrule
\end{tabular}
\label{tt4}
\end{table}

\begin{figure}
\centering
\includegraphics[width=7.4cm]{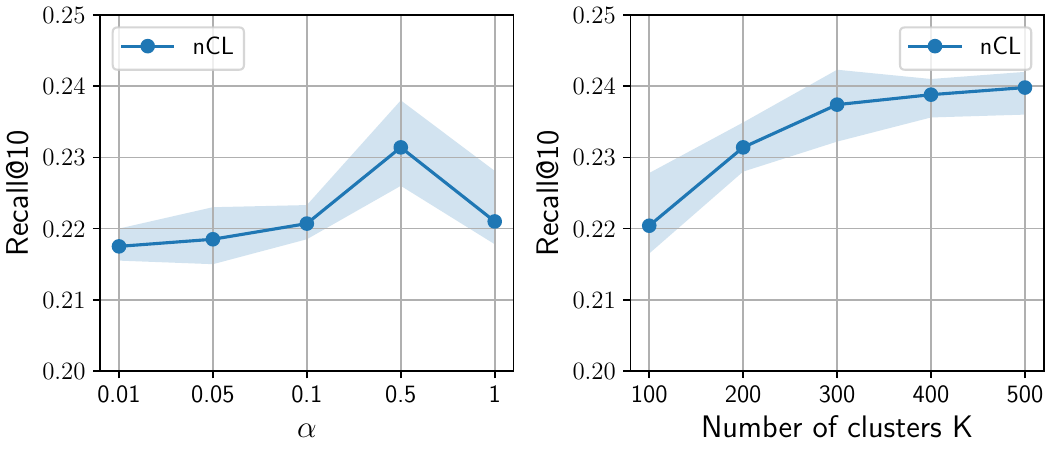}
\caption{ Results of parameters sensitivity ($\alpha$ and $K$) on MovieLens10M  dataset.
 }
\label{ss}
\end{figure}

\subsubsection{\textbf{Sensitive Analysis}}

Our proposed nCL has two major hyperparameters: $\alpha$, which balances the contributions of alignment and compactness, and $K$, which denotes the granularity of user/item cluster information. We vary the regularized parameter $\alpha$ in Eq. (\ref{eq16}) within the range of $\{0.01, 0.05, 0.1, 0.5, 1\}$ and the number of user/item clusters $K$ within the range of $\{100, 200, 300, 400, 500\}$.

Figure~\ref{ss} illustrates the impact of changing one parameter while fixing the other on MovieLens10M dataset. Our nCL algorithm shows relative stability across different parameter settings. Specifically, setting $\alpha = 0.5$ achieves a good balance between alignment and compactness of the embedding space, resulting in the best performance. For the number of clusters $K$, we find that the performance improves with larger $K$. This is expected since a higher number of clusters leads to more discriminative representations of users/items. In practice, around 300 clusters are sufficient to capture community information.

\section{Conclusion and Future Work}
In this study, we provide empirical evidence that existing contrastive methods are prone to the problem of dimensional collapse. To address this issue, we propose nCL, a simple yet effective method for collaborative filtering. Our nCL improves user/item representations by considering both the alignment and compactness of the embedding space. Extensive experiments demonstrate that nCL leads to better performance. As part of future work, we plan to explore and apply our non-contrastive loss function on various real-world applications such as sequential recommendation and fairness recommendation. We are also interested in further accelerating the computation of compactness by using variational functions.

\bibliographystyle{ACM-Reference-Format}
\bibliography{my}


\begin{thebibliography}{54}


\ifx \showCODEN    \undefined \def \showCODEN     #1{\unskip}     \fi
\ifx \showDOI      \undefined \def \showDOI       #1{#1}\fi
\ifx \showISBNx    \undefined \def \showISBNx     #1{\unskip}     \fi
\ifx \showISBNxiii \undefined \def \showISBNxiii  #1{\unskip}     \fi
\ifx \showISSN     \undefined \def \showISSN      #1{\unskip}     \fi
\ifx \showLCCN     \undefined \def \showLCCN      #1{\unskip}     \fi
\ifx \shownote     \undefined \def \shownote      #1{#1}          \fi
\ifx \showarticletitle \undefined \def \showarticletitle #1{#1}   \fi
\ifx \showURL      \undefined \def \showURL       {\relax}        \fi
\providecommand\bibfield[2]{#2}
\providecommand\bibinfo[2]{#2}
\providecommand\natexlab[1]{#1}
\providecommand\showeprint[2][]{arXiv:#2}

\bibitem[Asano et~al\mbox{.}(2019)]%
        {asano2019self}
\bibfield{author}{\bibinfo{person}{YM Asano}, \bibinfo{person}{C Rupprecht}, {and} \bibinfo{person}{A Vedaldi}.} \bibinfo{year}{2019}\natexlab{}.
\newblock \showarticletitle{Self-labelling via simultaneous clustering and representation learning}. In \bibinfo{booktitle}{\emph{International Conference on Learning Representations}}.
\newblock


\bibitem[Baek et~al\mbox{.}(2022)]%
        {baek2022efficient}
\bibfield{author}{\bibinfo{person}{Christina Baek}, \bibinfo{person}{Ziyang Wu}, \bibinfo{person}{Kwan Ho~Ryan Chan}, \bibinfo{person}{Tianjiao Ding}, \bibinfo{person}{Yi Ma}, {and} \bibinfo{person}{Benjamin~D Haeffele}.} \bibinfo{year}{2022}\natexlab{}.
\newblock \showarticletitle{Efficient Maximal Coding Rate Reduction by Variational Forms}. In \bibinfo{booktitle}{\emph{Proceedings of the IEEE/CVF Conference on Computer Vision and Pattern Recognition}}. \bibinfo{pages}{500--508}.
\newblock


\bibitem[Caron et~al\mbox{.}(2020)]%
        {caron2020unsupervised}
\bibfield{author}{\bibinfo{person}{Mathilde Caron}, \bibinfo{person}{Ishan Misra}, \bibinfo{person}{Julien Mairal}, \bibinfo{person}{Priya Goyal}, \bibinfo{person}{Piotr Bojanowski}, {and} \bibinfo{person}{Armand Joulin}.} \bibinfo{year}{2020}\natexlab{}.
\newblock \showarticletitle{Unsupervised learning of visual features by contrasting cluster assignments}. In \bibinfo{booktitle}{\emph{Advances in Neural Information Processing Systems}}.
\newblock


\bibitem[Chan et~al\mbox{.}(2022)]%
        {chan2022redunet}
\bibfield{author}{\bibinfo{person}{Kwan Ho~Ryan Chan}, \bibinfo{person}{Yaodong Yu}, \bibinfo{person}{Chong You}, \bibinfo{person}{Haozhi Qi}, \bibinfo{person}{John Wright}, {and} \bibinfo{person}{Yi Ma}.} \bibinfo{year}{2022}\natexlab{}.
\newblock \showarticletitle{ReduNet: A White-box Deep Network from the Principle of Maximizing Rate Reduction}.
\newblock \bibinfo{journal}{\emph{Journal of Machine Learning Research}} (\bibinfo{year}{2022}), \bibinfo{pages}{1--103}.
\newblock


\bibitem[Chen et~al\mbox{.}(2022a)]%
        {chen2022denoising}
\bibfield{author}{\bibinfo{person}{Huiyuan Chen}, \bibinfo{person}{Yusan Lin}, \bibinfo{person}{Menghai Pan}, \bibinfo{person}{Lan Wang}, \bibinfo{person}{Chin-Chia~Michael Yeh}, \bibinfo{person}{Xiaoting Li}, \bibinfo{person}{Yan Zheng}, \bibinfo{person}{Fei Wang}, {and} \bibinfo{person}{Hao Yang}.} \bibinfo{year}{2022}\natexlab{a}.
\newblock \showarticletitle{Denoising self-attentive sequential recommendation}. In \bibinfo{booktitle}{\emph{Proceedings of the 16th ACM Conference on Recommender Systems}}. \bibinfo{pages}{92--101}.
\newblock


\bibitem[Chen et~al\mbox{.}(2021)]%
        {chen2021structured}
\bibfield{author}{\bibinfo{person}{Huiyuan Chen}, \bibinfo{person}{Lan Wang}, \bibinfo{person}{Yusan Lin}, \bibinfo{person}{Chin-Chia~Michael Yeh}, \bibinfo{person}{Fei Wang}, {and} \bibinfo{person}{Hao Yang}.} \bibinfo{year}{2021}\natexlab{}.
\newblock \showarticletitle{Structured graph convolutional networks with stochastic masks for recommender systems}. In \bibinfo{booktitle}{\emph{Proceedings of the 44th International ACM SIGIR Conference on Research and Development in Information Retrieval}}. \bibinfo{pages}{614--623}.
\newblock


\bibitem[Chen et~al\mbox{.}(2023)]%
        {chen2023sharpness}
\bibfield{author}{\bibinfo{person}{Huiyuan Chen}, \bibinfo{person}{Chin-Chia~Michael Yeh}, \bibinfo{person}{Yujie Fan}, \bibinfo{person}{Yan Zheng}, \bibinfo{person}{Junpeng Wang}, \bibinfo{person}{Vivian Lai}, \bibinfo{person}{Mahashweta Das}, {and} \bibinfo{person}{Hao Yang}.} \bibinfo{year}{2023}\natexlab{}.
\newblock \showarticletitle{Sharpness-Aware Graph Collaborative Filtering}. In \bibinfo{booktitle}{\emph{Proceedings of the 46th International ACM SIGIR Conference on Research and Development in Information Retrieval}}. \bibinfo{pages}{2369--2373}.
\newblock


\bibitem[Chen et~al\mbox{.}(2022b)]%
        {chen2022graph}
\bibfield{author}{\bibinfo{person}{Huiyuan Chen}, \bibinfo{person}{Chin-Chia~Michael Yeh}, \bibinfo{person}{Fei Wang}, {and} \bibinfo{person}{Hao Yang}.} \bibinfo{year}{2022}\natexlab{b}.
\newblock \showarticletitle{Graph neural transport networks with non-local attentions for recommender systems}. In \bibinfo{booktitle}{\emph{Proceedings of the ACM Web Conference 2022}}. \bibinfo{pages}{1955--1964}.
\newblock


\bibitem[Chen et~al\mbox{.}(2020)]%
        {chen2020simple}
\bibfield{author}{\bibinfo{person}{Ting Chen}, \bibinfo{person}{Simon Kornblith}, \bibinfo{person}{Mohammad Norouzi}, {and} \bibinfo{person}{Geoffrey Hinton}.} \bibinfo{year}{2020}\natexlab{}.
\newblock \showarticletitle{A simple framework for contrastive learning of visual representations}. In \bibinfo{booktitle}{\emph{International Conference on Machine Learning}}. \bibinfo{pages}{1597--1607}.
\newblock


\bibitem[Cuturi(2013)]%
        {cuturi2013sinkhorn}
\bibfield{author}{\bibinfo{person}{Marco Cuturi}.} \bibinfo{year}{2013}\natexlab{}.
\newblock \showarticletitle{Sinkhorn distances: Lightspeed computation of optimal transport}.
\newblock \bibinfo{journal}{\emph{Advances in neural information processing systems}} (\bibinfo{year}{2013}).
\newblock


\bibitem[Gao et~al\mbox{.}(2021)]%
        {gao2021simcse}
\bibfield{author}{\bibinfo{person}{Tianyu Gao}, \bibinfo{person}{Xingcheng Yao}, {and} \bibinfo{person}{Danqi Chen}.} \bibinfo{year}{2021}\natexlab{}.
\newblock \showarticletitle{SimCSE: Simple Contrastive Learning of Sentence Embeddings}. In \bibinfo{booktitle}{\emph{Proceedings of the 2021 Conference on Empirical Methods in Natural Language Processing}}. \bibinfo{pages}{6894--6910}.
\newblock


\bibitem[Guo et~al\mbox{.}(2017)]%
        {guo2017deepfm}
\bibfield{author}{\bibinfo{person}{Huifeng Guo}, \bibinfo{person}{Ruiming Tang}, \bibinfo{person}{Yunming Ye}, \bibinfo{person}{Zhenguo Li}, {and} \bibinfo{person}{Xiuqiang He}.} \bibinfo{year}{2017}\natexlab{}.
\newblock \showarticletitle{DeepFM: a factorization-machine based neural network for CTR prediction}.
\newblock \bibinfo{journal}{\emph{arXiv preprint arXiv:1703.04247}} (\bibinfo{year}{2017}).
\newblock


\bibitem[Han et~al\mbox{.}(2022)]%
        {han2022geometric}
\bibfield{author}{\bibinfo{person}{Xiaotian Han}, \bibinfo{person}{Zhimeng Jiang}, \bibinfo{person}{Ninghao Liu}, \bibinfo{person}{Qingquan Song}, \bibinfo{person}{Jundong Li}, {and} \bibinfo{person}{Xia Hu}.} \bibinfo{year}{2022}\natexlab{}.
\newblock \showarticletitle{Geometric Graph Representation Learning via Maximizing Rate Reduction}. In \bibinfo{booktitle}{\emph{Proceedings of the ACM Web Conference 2022}}. \bibinfo{pages}{1226--1237}.
\newblock


\bibitem[He and Ozay(2022)]%
        {he2022exploring}
\bibfield{author}{\bibinfo{person}{Bobby He} {and} \bibinfo{person}{Mete Ozay}.} \bibinfo{year}{2022}\natexlab{}.
\newblock \showarticletitle{Exploring the Gap between Collapsed \& Whitened Features in Self-Supervised Learning}. In \bibinfo{booktitle}{\emph{International Conference on Machine Learning}}. \bibinfo{pages}{8613--8634}.
\newblock


\bibitem[He et~al\mbox{.}(2020)]%
        {he2020lightgcn}
\bibfield{author}{\bibinfo{person}{Xiangnan He}, \bibinfo{person}{Kuan Deng}, \bibinfo{person}{Xiang Wang}, \bibinfo{person}{Yan Li}, \bibinfo{person}{Yongdong Zhang}, {and} \bibinfo{person}{Meng Wang}.} \bibinfo{year}{2020}\natexlab{}.
\newblock \showarticletitle{Lightgcn: Simplifying and powering graph convolution network for recommendation}. In \bibinfo{booktitle}{\emph{Proceedings of the 43rd International ACM SIGIR conference on research and development in Information Retrieval}}. \bibinfo{pages}{639--648}.
\newblock


\bibitem[He et~al\mbox{.}(2017)]%
        {he2017neural}
\bibfield{author}{\bibinfo{person}{Xiangnan He}, \bibinfo{person}{Lizi Liao}, \bibinfo{person}{Hanwang Zhang}, \bibinfo{person}{Liqiang Nie}, \bibinfo{person}{Xia Hu}, {and} \bibinfo{person}{Tat-Seng Chua}.} \bibinfo{year}{2017}\natexlab{}.
\newblock \showarticletitle{Neural collaborative filtering}. In \bibinfo{booktitle}{\emph{Proceedings of the 26th International Conference on World Wide Web}}. International World Wide Web Conferences Steering Committee, \bibinfo{pages}{173--182}.
\newblock


\bibitem[Hua et~al\mbox{.}(2021)]%
        {hua2021feature}
\bibfield{author}{\bibinfo{person}{Tianyu Hua}, \bibinfo{person}{Wenxiao Wang}, \bibinfo{person}{Zihui Xue}, \bibinfo{person}{Sucheng Ren}, \bibinfo{person}{Yue Wang}, {and} \bibinfo{person}{Hang Zhao}.} \bibinfo{year}{2021}\natexlab{}.
\newblock \showarticletitle{On feature decorrelation in self-supervised learning}. In \bibinfo{booktitle}{\emph{Proceedings of the IEEE/CVF International Conference on Computer Vision}}. \bibinfo{pages}{9598--9608}.
\newblock


\bibitem[Jing et~al\mbox{.}(2021)]%
        {jing2021understanding}
\bibfield{author}{\bibinfo{person}{Li Jing}, \bibinfo{person}{Pascal Vincent}, \bibinfo{person}{Yann LeCun}, {and} \bibinfo{person}{Yuandong Tian}.} \bibinfo{year}{2021}\natexlab{}.
\newblock \showarticletitle{Understanding Dimensional Collapse in Contrastive Self-supervised Learning}. In \bibinfo{booktitle}{\emph{International Conference on Learning Representations}}.
\newblock


\bibitem[Koren et~al\mbox{.}(2009)]%
        {koren2009matrix}
\bibfield{author}{\bibinfo{person}{Yehuda Koren}, \bibinfo{person}{Robert Bell}, {and} \bibinfo{person}{Chris Volinsky}.} \bibinfo{year}{2009}\natexlab{}.
\newblock \showarticletitle{Matrix factorization techniques for recommender systems}.
\newblock \bibinfo{journal}{\emph{Computer}} \bibinfo{volume}{42}, \bibinfo{number}{8} (\bibinfo{year}{2009}), \bibinfo{pages}{30--37}.
\newblock


\bibitem[Krichene and Rendle(2020)]%
        {DBLP:conf/kdd/KricheneR20}
\bibfield{author}{\bibinfo{person}{Walid Krichene} {and} \bibinfo{person}{Steffen Rendle}.} \bibinfo{year}{2020}\natexlab{}.
\newblock \showarticletitle{On sampled metrics for item recommendation}. In \bibinfo{booktitle}{\emph{Proceedings of the 26th ACM SIGKDD International Conference on Knowledge Discovery \& Data Mining}}. \bibinfo{pages}{1748--1757}.
\newblock


\bibitem[Lai et~al\mbox{.}(2023)]%
        {lai2023enhancing}
\bibfield{author}{\bibinfo{person}{Vivian Lai}, \bibinfo{person}{Huiyuan Chen}, \bibinfo{person}{Chin-Chia~Michael Yeh}, \bibinfo{person}{Minghua Xu}, \bibinfo{person}{Yiwei Cai}, {and} \bibinfo{person}{Hao Yang}.} \bibinfo{year}{2023}\natexlab{}.
\newblock \showarticletitle{Enhancing Transformers without Self-supervised Learning: A Loss Landscape Perspective in Sequential Recommendation}. In \bibinfo{booktitle}{\emph{Proceedings of the 17th ACM Conference on Recommender Systems}}. \bibinfo{pages}{791--797}.
\newblock


\bibitem[Lee et~al\mbox{.}(2021)]%
        {lee2021bootstrapping}
\bibfield{author}{\bibinfo{person}{Dongha Lee}, \bibinfo{person}{SeongKu Kang}, \bibinfo{person}{Hyunjun Ju}, \bibinfo{person}{Chanyoung Park}, {and} \bibinfo{person}{Hwanjo Yu}.} \bibinfo{year}{2021}\natexlab{}.
\newblock \showarticletitle{Bootstrapping user and item representations for one-class collaborative filtering}. In \bibinfo{booktitle}{\emph{Proceedings of the 44th International ACM SIGIR Conference on Research and Development in Information Retrieval}}. \bibinfo{pages}{317--326}.
\newblock


\bibitem[Liang et~al\mbox{.}(2023a)]%
        {liang2023knowledge}
\bibfield{author}{\bibinfo{person}{Ke Liang}, \bibinfo{person}{Yue Liu}, \bibinfo{person}{Sihang Zhou}, \bibinfo{person}{Wenxuan Tu}, \bibinfo{person}{Yi Wen}, \bibinfo{person}{Xihong Yang}, \bibinfo{person}{Xiangjun Dong}, {and} \bibinfo{person}{Xinwang Liu}.} \bibinfo{year}{2023}\natexlab{a}.
\newblock \showarticletitle{Knowledge Graph Contrastive Learning Based on Relation-Symmetrical Structure}.
\newblock \bibinfo{journal}{\emph{IEEE Transactions on Knowledge and Data Engineering}} (\bibinfo{year}{2023}).
\newblock


\bibitem[Liang et~al\mbox{.}(2023b)]%
        {liang2023learn}
\bibfield{author}{\bibinfo{person}{Ke Liang}, \bibinfo{person}{Lingyuan Meng}, \bibinfo{person}{Meng Liu}, \bibinfo{person}{Yue Liu}, \bibinfo{person}{Wenxuan Tu}, \bibinfo{person}{Siwei Wang}, \bibinfo{person}{Sihang Zhou}, {and} \bibinfo{person}{Xinwang Liu}.} \bibinfo{year}{2023}\natexlab{b}.
\newblock \showarticletitle{Learn from relational correlations and periodic events for temporal knowledge graph reasoning}. In \bibinfo{booktitle}{\emph{Proceedings of the 46th International ACM SIGIR Conference on Research and Development in Information Retrieval}}. \bibinfo{pages}{1559--1568}.
\newblock


\bibitem[Liang et~al\mbox{.}(2022)]%
        {liang2022reasoning}
\bibfield{author}{\bibinfo{person}{Ke Liang}, \bibinfo{person}{Lingyuan Meng}, \bibinfo{person}{Meng Liu}, \bibinfo{person}{Yue Liu}, \bibinfo{person}{Wenxuan Tu}, \bibinfo{person}{Siwei Wang}, \bibinfo{person}{Sihang Zhou}, \bibinfo{person}{Xinwang Liu}, {and} \bibinfo{person}{Fuchun Sun}.} \bibinfo{year}{2022}\natexlab{}.
\newblock \showarticletitle{Reasoning over different types of knowledge graphs: Static, temporal and multi-modal}.
\newblock \bibinfo{journal}{\emph{arXiv preprint arXiv:2212.05767}} (\bibinfo{year}{2022}).
\newblock


\bibitem[Lin et~al\mbox{.}(2022)]%
        {lin2022improving}
\bibfield{author}{\bibinfo{person}{Zihan Lin}, \bibinfo{person}{Changxin Tian}, \bibinfo{person}{Yupeng Hou}, {and} \bibinfo{person}{Wayne~Xin Zhao}.} \bibinfo{year}{2022}\natexlab{}.
\newblock \showarticletitle{Improving Graph Collaborative Filtering with Neighborhood-enriched Contrastive Learning}. In \bibinfo{booktitle}{\emph{Proceedings of the ACM Web Conference 2022}}. \bibinfo{pages}{2320--2329}.
\newblock


\bibitem[Luo et~al\mbox{.}(2020)]%
        {luo2020parameterized}
\bibfield{author}{\bibinfo{person}{Dongsheng Luo}, \bibinfo{person}{Wei Cheng}, \bibinfo{person}{Dongkuan Xu}, \bibinfo{person}{Wenchao Yu}, \bibinfo{person}{Bo Zong}, \bibinfo{person}{Haifeng Chen}, {and} \bibinfo{person}{Xiang Zhang}.} \bibinfo{year}{2020}\natexlab{}.
\newblock \showarticletitle{Parameterized explainer for graph neural network}.
\newblock \bibinfo{journal}{\emph{Advances in neural information processing systems}}  \bibinfo{volume}{33} (\bibinfo{year}{2020}), \bibinfo{pages}{19620--19631}.
\newblock


\bibitem[Luo et~al\mbox{.}(2021)]%
        {luo2021learning}
\bibfield{author}{\bibinfo{person}{Dongsheng Luo}, \bibinfo{person}{Wei Cheng}, \bibinfo{person}{Wenchao Yu}, \bibinfo{person}{Bo Zong}, \bibinfo{person}{Jingchao Ni}, \bibinfo{person}{Haifeng Chen}, {and} \bibinfo{person}{Xiang Zhang}.} \bibinfo{year}{2021}\natexlab{}.
\newblock \showarticletitle{Learning to drop: Robust graph neural network via topological denoising}. In \bibinfo{booktitle}{\emph{Proceedings of the 14th ACM international conference on web search and data mining}}. \bibinfo{pages}{779--787}.
\newblock


\bibitem[Mao et~al\mbox{.}(2021a)]%
        {mao2021simplex}
\bibfield{author}{\bibinfo{person}{Kelong Mao}, \bibinfo{person}{Jieming Zhu}, \bibinfo{person}{Jinpeng Wang}, \bibinfo{person}{Quanyu Dai}, \bibinfo{person}{Zhenhua Dong}, \bibinfo{person}{Xi Xiao}, {and} \bibinfo{person}{Xiuqiang He}.} \bibinfo{year}{2021}\natexlab{a}.
\newblock \showarticletitle{SimpleX: A Simple and Strong Baseline for Collaborative Filtering}. In \bibinfo{booktitle}{\emph{Proceedings of the 30th ACM International Conference on Information \& Knowledge Management}}. \bibinfo{pages}{1243--1252}.
\newblock


\bibitem[Mao et~al\mbox{.}(2021b)]%
        {mao2021ultragcn}
\bibfield{author}{\bibinfo{person}{Kelong Mao}, \bibinfo{person}{Jieming Zhu}, \bibinfo{person}{Xi Xiao}, \bibinfo{person}{Biao Lu}, \bibinfo{person}{Zhaowei Wang}, {and} \bibinfo{person}{Xiuqiang He}.} \bibinfo{year}{2021}\natexlab{b}.
\newblock \showarticletitle{UltraGCN: ultra simplification of graph convolutional networks for recommendation}. In \bibinfo{booktitle}{\emph{Proceedings of the 30th ACM International Conference on Information \& Knowledge Management}}. \bibinfo{pages}{1253--1262}.
\newblock


\bibitem[Monti et~al\mbox{.}(2017)]%
        {monti2017geometric}
\bibfield{author}{\bibinfo{person}{Federico Monti}, \bibinfo{person}{Michael Bronstein}, {and} \bibinfo{person}{Xavier Bresson}.} \bibinfo{year}{2017}\natexlab{}.
\newblock \showarticletitle{Geometric matrix completion with recurrent multi-graph neural networks}.
\newblock \bibinfo{journal}{\emph{Advances in neural information processing systems}}  \bibinfo{volume}{30} (\bibinfo{year}{2017}).
\newblock


\bibitem[Oord et~al\mbox{.}(2018)]%
        {oord2018representation}
\bibfield{author}{\bibinfo{person}{Aaron van~den Oord}, \bibinfo{person}{Yazhe Li}, {and} \bibinfo{person}{Oriol Vinyals}.} \bibinfo{year}{2018}\natexlab{}.
\newblock \showarticletitle{Representation learning with contrastive predictive coding}.
\newblock \bibinfo{journal}{\emph{arXiv preprint arXiv:1807.03748}} (\bibinfo{year}{2018}).
\newblock


\bibitem[Rendle et~al\mbox{.}(2009)]%
        {rendle2009bpr}
\bibfield{author}{\bibinfo{person}{Steffen Rendle}, \bibinfo{person}{Christoph Freudenthaler}, \bibinfo{person}{Zeno Gantner}, {and} \bibinfo{person}{Lars Schmidt-Thieme}.} \bibinfo{year}{2009}\natexlab{}.
\newblock \showarticletitle{BPR: Bayesian personalized ranking from implicit feedback}. In \bibinfo{booktitle}{\emph{Proceedings of the 25th conference on uncertainty in artificial intelligence}}. \bibinfo{pages}{452--461}.
\newblock


\bibitem[Thomas and Joy(2006)]%
        {thomas2006elements}
\bibfield{author}{\bibinfo{person}{Mtcaj Thomas} {and} \bibinfo{person}{A~Thomas Joy}.} \bibinfo{year}{2006}\natexlab{}.
\newblock \bibinfo{booktitle}{\emph{Elements of information theory}}.
\newblock \bibinfo{publisher}{Wiley-Interscience}.
\newblock


\bibitem[Wang et~al\mbox{.}(2022a)]%
        {wang2022towards}
\bibfield{author}{\bibinfo{person}{Chenyang Wang}, \bibinfo{person}{Yuanqing Yu}, \bibinfo{person}{Weizhi Ma}, \bibinfo{person}{Min Zhang}, \bibinfo{person}{Chong Chen}, \bibinfo{person}{Yiqun Liu}, {and} \bibinfo{person}{Shaoping Ma}.} \bibinfo{year}{2022}\natexlab{a}.
\newblock \showarticletitle{Towards Representation Alignment and Uniformity in Collaborative Filtering}. In \bibinfo{booktitle}{\emph{Proceedings of the 28th ACM SIGKDD Conference on Knowledge Discovery and Data Mining}}. \bibinfo{pages}{1816--1825}.
\newblock


\bibitem[Wang et~al\mbox{.}(2023)]%
        {song2023}
\bibfield{author}{\bibinfo{person}{Song Wang}, \bibinfo{person}{Xingbo Fu}, \bibinfo{person}{Kaize Ding}, \bibinfo{person}{Chen Chen}, \bibinfo{person}{Huiyuan Chen}, {and} \bibinfo{person}{Jundong Li}.} \bibinfo{year}{2023}\natexlab{}.
\newblock \showarticletitle{Federated Few-Shot Learning}. In \bibinfo{booktitle}{\emph{Proceedings of the 29th ACM SIGKDD Conference on Knowledge Discovery and Data Mining}}. \bibinfo{pages}{2374–2385}.
\newblock


\bibitem[Wang and Isola(2020)]%
        {wang2020understanding}
\bibfield{author}{\bibinfo{person}{Tongzhou Wang} {and} \bibinfo{person}{Phillip Isola}.} \bibinfo{year}{2020}\natexlab{}.
\newblock \showarticletitle{Understanding contrastive representation learning through alignment and uniformity on the hypersphere}. In \bibinfo{booktitle}{\emph{International Conference on Machine Learning}}. \bibinfo{pages}{9929--9939}.
\newblock


\bibitem[Wang et~al\mbox{.}(2019)]%
        {wang2019neural}
\bibfield{author}{\bibinfo{person}{Xiang Wang}, \bibinfo{person}{Xiangnan He}, \bibinfo{person}{Meng Wang}, \bibinfo{person}{Fuli Feng}, {and} \bibinfo{person}{Tat-Seng Chua}.} \bibinfo{year}{2019}\natexlab{}.
\newblock \showarticletitle{Neural graph collaborative filtering}. In \bibinfo{booktitle}{\emph{Proceedings of the 42nd international ACM SIGIR conference on Research and development in Information Retrieval}}. \bibinfo{pages}{165--174}.
\newblock


\bibitem[Wang et~al\mbox{.}(2020)]%
        {wang2020disentangled}
\bibfield{author}{\bibinfo{person}{Xiang Wang}, \bibinfo{person}{Hongye Jin}, \bibinfo{person}{An Zhang}, \bibinfo{person}{Xiangnan He}, \bibinfo{person}{Tong Xu}, {and} \bibinfo{person}{Tat-Seng Chua}.} \bibinfo{year}{2020}\natexlab{}.
\newblock \showarticletitle{Disentangled graph collaborative filtering}. In \bibinfo{booktitle}{\emph{Proceedings of the 43rd International ACM SIGIR Conference on Research and Development in Information Retrieval}}. \bibinfo{pages}{1001--1010}.
\newblock


\bibitem[Wang et~al\mbox{.}(2022b)]%
        {wang2022clusterscl}
\bibfield{author}{\bibinfo{person}{Yanling Wang}, \bibinfo{person}{Jing Zhang}, \bibinfo{person}{Haoyang Li}, \bibinfo{person}{Yuxiao Dong}, \bibinfo{person}{Hongzhi Yin}, \bibinfo{person}{Cuiping Li}, {and} \bibinfo{person}{Hong Chen}.} \bibinfo{year}{2022}\natexlab{b}.
\newblock \showarticletitle{ClusterSCL: Cluster-Aware Supervised Contrastive Learning on Graphs}. In \bibinfo{booktitle}{\emph{Proceedings of the ACM Web Conference 2022}}. \bibinfo{pages}{1611--1621}.
\newblock


\bibitem[Wang et~al\mbox{.}(2022c)]%
        {wang2022improving}
\bibfield{author}{\bibinfo{person}{Yu Wang}, \bibinfo{person}{Yuying Zhao}, \bibinfo{person}{Yushun Dong}, \bibinfo{person}{Huiyuan Chen}, \bibinfo{person}{Jundong Li}, {and} \bibinfo{person}{Tyler Derr}.} \bibinfo{year}{2022}\natexlab{c}.
\newblock \showarticletitle{Improving fairness in graph neural networks via mitigating sensitive attribute leakage}. In \bibinfo{booktitle}{\emph{Proceedings of the 28th ACM SIGKDD Conference on Knowledge Discovery and Data Mining}}. \bibinfo{pages}{1938--1948}.
\newblock


\bibitem[Wu et~al\mbox{.}(2021)]%
        {wu2021self}
\bibfield{author}{\bibinfo{person}{Jiancan Wu}, \bibinfo{person}{Xiang Wang}, \bibinfo{person}{Fuli Feng}, \bibinfo{person}{Xiangnan He}, \bibinfo{person}{Liang Chen}, \bibinfo{person}{Jianxun Lian}, {and} \bibinfo{person}{Xing Xie}.} \bibinfo{year}{2021}\natexlab{}.
\newblock \showarticletitle{Self-supervised graph learning for recommendation}. In \bibinfo{booktitle}{\emph{Proceedings of the 44th International ACM SIGIR Conference on Research and Development in Information Retrieval}}. \bibinfo{pages}{726--735}.
\newblock


\bibitem[Xie et~al\mbox{.}(2020)]%
        {xie2020fast}
\bibfield{author}{\bibinfo{person}{Yujia Xie}, \bibinfo{person}{Xiangfeng Wang}, \bibinfo{person}{Ruijia Wang}, {and} \bibinfo{person}{Hongyuan Zha}.} \bibinfo{year}{2020}\natexlab{}.
\newblock \showarticletitle{A fast proximal point method for computing exact wasserstein distance}. In \bibinfo{booktitle}{\emph{Uncertainty in artificial intelligence}}. \bibinfo{pages}{433--453}.
\newblock


\bibitem[Xu et~al\mbox{.}(2023)]%
        {xu2023kernel}
\bibfield{author}{\bibinfo{person}{Zhe Xu}, \bibinfo{person}{Yuzhong Chen}, \bibinfo{person}{Menghai Pan}, \bibinfo{person}{Huiyuan Chen}, \bibinfo{person}{Mahashweta Das}, \bibinfo{person}{Hao Yang}, {and} \bibinfo{person}{Hanghang Tong}.} \bibinfo{year}{2023}\natexlab{}.
\newblock \showarticletitle{Kernel Ridge Regression-Based Graph Dataset Distillation}. In \bibinfo{booktitle}{\emph{Proceedings of the 29th ACM SIGKDD Conference on Knowledge Discovery and Data Mining}}. \bibinfo{pages}{2850--2861}.
\newblock


\bibitem[Yan et~al\mbox{.}(2023)]%
        {yan2023from}
\bibfield{author}{\bibinfo{person}{Yuchen Yan}, \bibinfo{person}{Yuzhong Chen}, \bibinfo{person}{Huiyuan Chen}, \bibinfo{person}{Minghua Xu}, \bibinfo{person}{Mahashweta Das}, \bibinfo{person}{Hao Yang}, {and} \bibinfo{person}{Hanghang Tong}.} \bibinfo{year}{2023}\natexlab{}.
\newblock \showarticletitle{From Trainable Negative Depth to Edge Heterophily in Graphs}. In \bibinfo{booktitle}{\emph{Thirty-seventh Conference on Neural Information Processing Systems}}.
\newblock


\bibitem[Yeh et~al\mbox{.}(2022)]%
        {yeh2022embedding}
\bibfield{author}{\bibinfo{person}{Chin-Chia~Michael Yeh}, \bibinfo{person}{Mengting Gu}, \bibinfo{person}{Yan Zheng}, \bibinfo{person}{Huiyuan Chen}, \bibinfo{person}{Javid Ebrahimi}, \bibinfo{person}{Zhongfang Zhuang}, \bibinfo{person}{Junpeng Wang}, \bibinfo{person}{Liang Wang}, {and} \bibinfo{person}{Wei Zhang}.} \bibinfo{year}{2022}\natexlab{}.
\newblock \showarticletitle{Embedding Compression with Hashing for Efficient Representation Learning in Large-Scale Graph}. In \bibinfo{booktitle}{\emph{Proceedings of the 28th ACM SIGKDD Conference on Knowledge Discovery and Data Mining}}. \bibinfo{pages}{4391--4401}.
\newblock


\bibitem[You et~al\mbox{.}(2020)]%
        {you2020graph}
\bibfield{author}{\bibinfo{person}{Yuning You}, \bibinfo{person}{Tianlong Chen}, \bibinfo{person}{Yongduo Sui}, \bibinfo{person}{Ting Chen}, \bibinfo{person}{Zhangyang Wang}, {and} \bibinfo{person}{Yang Shen}.} \bibinfo{year}{2020}\natexlab{}.
\newblock \showarticletitle{Graph contrastive learning with augmentations}.
\newblock \bibinfo{journal}{\emph{Advances in Neural Information Processing Systems}} (\bibinfo{year}{2020}), \bibinfo{pages}{5812--5823}.
\newblock


\bibitem[Yu et~al\mbox{.}(2022)]%
        {yu2022graph}
\bibfield{author}{\bibinfo{person}{Junliang Yu}, \bibinfo{person}{Hongzhi Yin}, \bibinfo{person}{Xin Xia}, \bibinfo{person}{Tong Chen}, \bibinfo{person}{Lizhen Cui}, {and} \bibinfo{person}{Quoc Viet~Hung Nguyen}.} \bibinfo{year}{2022}\natexlab{}.
\newblock \showarticletitle{Are graph augmentations necessary? simple graph contrastive learning for recommendation}. In \bibinfo{booktitle}{\emph{Proceedings of the 45th International ACM SIGIR Conference on Research and Development in Information Retrieval}}. \bibinfo{pages}{1294--1303}.
\newblock


\bibitem[Yu et~al\mbox{.}(2020)]%
        {yu2020learning}
\bibfield{author}{\bibinfo{person}{Yaodong Yu}, \bibinfo{person}{Kwan Ho~Ryan Chan}, \bibinfo{person}{Chong You}, \bibinfo{person}{Chaobing Song}, {and} \bibinfo{person}{Yi Ma}.} \bibinfo{year}{2020}\natexlab{}.
\newblock \showarticletitle{Learning diverse and discriminative representations via the principle of maximal coding rate reduction}.
\newblock \bibinfo{journal}{\emph{Advances in Neural Information Processing Systems}} (\bibinfo{year}{2020}), \bibinfo{pages}{9422--9434}.
\newblock


\bibitem[Zhang et~al\mbox{.}(2021)]%
        {zhang2021universal}
\bibfield{author}{\bibinfo{person}{George Zhang}, \bibinfo{person}{Jingjing Qian}, \bibinfo{person}{Jun Chen}, {and} \bibinfo{person}{Ashish Khisti}.} \bibinfo{year}{2021}\natexlab{}.
\newblock \showarticletitle{Universal rate-distortion-perception representations for lossy compression}.
\newblock \bibinfo{journal}{\emph{Advances in Neural Information Processing Systems}} (\bibinfo{year}{2021}), \bibinfo{pages}{11517--11529}.
\newblock


\bibitem[Zhang et~al\mbox{.}(2023)]%
        {zhang2023mixupexplainer}
\bibfield{author}{\bibinfo{person}{Jiaxing Zhang}, \bibinfo{person}{Dongsheng Luo}, {and} \bibinfo{person}{Hua Wei}.} \bibinfo{year}{2023}\natexlab{}.
\newblock \showarticletitle{MixupExplainer: Generalizing Explanations for Graph Neural Networks with Data Augmentation}. In \bibinfo{booktitle}{\emph{Proceedings of the 29th ACM SIGKDD Conference on Knowledge Discovery and Data Mining}}. \bibinfo{pages}{3286--3296}.
\newblock


\bibitem[Zhou et~al\mbox{.}(2021)]%
        {zhou2021contrastive}
\bibfield{author}{\bibinfo{person}{Chang Zhou}, \bibinfo{person}{Jianxin Ma}, \bibinfo{person}{Jianwei Zhang}, \bibinfo{person}{Jingren Zhou}, {and} \bibinfo{person}{Hongxia Yang}.} \bibinfo{year}{2021}\natexlab{}.
\newblock \showarticletitle{Contrastive learning for debiased candidate generation in large-scale recommender systems}. In \bibinfo{booktitle}{\emph{Proceedings of the 27th ACM SIGKDD Conference on Knowledge Discovery \& Data Mining}}. \bibinfo{pages}{3985--3995}.
\newblock


\bibitem[Zhou et~al\mbox{.}(2018)]%
        {zhou2018deep}
\bibfield{author}{\bibinfo{person}{Guorui Zhou}, \bibinfo{person}{Xiaoqiang Zhu}, \bibinfo{person}{Chenru Song}, \bibinfo{person}{Ying Fan}, \bibinfo{person}{Han Zhu}, \bibinfo{person}{Xiao Ma}, \bibinfo{person}{Yanghui Yan}, \bibinfo{person}{Junqi Jin}, \bibinfo{person}{Han Li}, {and} \bibinfo{person}{Kun Gai}.} \bibinfo{year}{2018}\natexlab{}.
\newblock \showarticletitle{Deep interest network for click-through rate prediction}. In \bibinfo{booktitle}{\emph{Proceedings of the 24th ACM SIGKDD international conference on knowledge discovery \& data mining}}. \bibinfo{pages}{1059--1068}.
\newblock


\bibitem[Zhou et~al\mbox{.}(2020)]%
        {zhou2020s3}
\bibfield{author}{\bibinfo{person}{Kun Zhou}, \bibinfo{person}{Hui Wang}, \bibinfo{person}{Wayne~Xin Zhao}, \bibinfo{person}{Yutao Zhu}, \bibinfo{person}{Sirui Wang}, \bibinfo{person}{Fuzheng Zhang}, \bibinfo{person}{Zhongyuan Wang}, {and} \bibinfo{person}{Ji-Rong Wen}.} \bibinfo{year}{2020}\natexlab{}.
\newblock \showarticletitle{S3-rec: Self-supervised learning for sequential recommendation with mutual information maximization}. In \bibinfo{booktitle}{\emph{Proceedings of the 29th ACM International Conference on Information \& Knowledge Management}}. \bibinfo{pages}{1893--1902}.
\newblock


\end{thebibliography}

\end{document}